# Label-free Imaging of Catalytic $H_2O_2$ Decomposition on Single Colloidal Pt Nanoparticles using Nanofluidic Scattering Microscopy

*Björn Altenburger[1], Carl Andersson[1], Sune Levin[2], Fredrik Westerlund[2], Joachim Fritzsche[1] and Christoph Langhammer[1]\**

[1]Department of Physics, Chalmers University of Technology; SE-412 96 Gothenburg, Sweden

[2]Department of Life Sciences, Chalmers University of Technology; SE-412 96 Gothenburg, Sweden

*Corresponding author: clangham@chalmers.se

**Abstract**

Single particle catalysis aims at determining factors that dictate nanoparticle activity and selectivity. Existing methods often use fluorescent model reactions at low reactant concentrations, operate at low pressures, or rely on plasmonic enhancement effects. Hence, methods to measure single nanoparticle activity at technically relevant conditions, and without fluorescence or other enhancement mechanisms, are still lacking. Here, we introduce nanofluidic scattering microscopy of catalytic reactions on single colloidal nanoparticles trapped inside nanofluidic channels to fill this gap. By detecting minuscule refractive index changes in a liquid flushed trough a nanochannel, we demonstrate that local $H_2O_2$ concentration changes in water can be accurately measured. Applying this principle, we analyze the $H_2O_2$ concentration profiles adjacent to single colloidal Pt nanoparticles during catalytic $H_2O_2$ decomposition into $O_2$ and $H_2O$ and derive the particles' individual turnover frequencies from the growth rate of $O_2$ gas bubbles formed in their respective nanochannel during reaction.

## Introduction

The vision of single particle catalysis is to directly correlate the activity or selectivity of a single nanoparticle obtained at practically relevant conditions with structural and chemical descriptors of that same particle. This is driven by the prospect of deeper fundamental insights, since catalytic properties of nanoparticles traditionally are evaluated at the ensemble level, that is, by averaging the response from a large number of them. From an atomistic perspective, however, this is problematic because nanoparticles are individuals in terms of their atomic arrangement and defects, and because they are dynamic entities in reaction conditions. Therefore, ensemble averaging carries the inherent risk of erroneous structure-function correlations and that, for example, the most selective or most active "champion" nanoparticle types are hidden in the average.

To date, several experimental approaches for the study of catalytic processes on single nanoparticles exist[1–11]. In brief, the reported methods rely on the sensitive detection of photon or electron signals that report on either the catalyst particle itself, on the product molecules formed, on reactant molecules consumed, or on temperature changes that the reaction evokes in the particle surrounding. However, none of these methods can provide direct single particle activity information without (i) either using fluorescence- based readout in a direct or indirect[12] manner that often limits the reaction conditions to ultralow concentrations and that cannot be applied to technically relevant reactions or (ii) using plasmonic enhancement effects when tip-enhanced Raman spectroscopy (TERS) is used, as it is also done for larger ensembles with for example surface enhanced Raman spectroscopy (SERS) and shell-isolated nanoparticle-enhanced Raman spectroscopy (SHINERS). Techniques that utilize electron microscopy approaches can reveal atomistic processes and changes on single particles via, for example, field emission spectroscopy (FEM)[13] but require often near-vacuum conditions. Even though TEM-approaches have been developed that can monitor particles in-situ[14–17], the required setup and fluidic chips are highly complex and do not directly resolve the activity of a single particle.

To contribute to the quest of experimental method development that enables efficient and quantitative scrutiny of catalytic reactions on single nanoparticles, we have recently introduced the concept of nanofluidic reactors and used them in combination with plasmonic imaging and spectroscopy, together with mass spectrometry in the gas phase[18–21], and with fluorescence microscopy in the liquid phase, using both nanofabricated particles and individually trapped colloidal nanocrystals as the catalyst[22,23]. One of the key traits of this nanofluidic reactor platform is that it ensures identical reaction conditions for the individual particles during an experiment while isolating each of them in its own nanochannel. Thereby, it enables highly parallelized studies of tens of single nanoparticles in the same experiment, whereby it eliminates errors and uncertainties inherent to subsequent experiments traditionally used. In our first studies, this approach has made it possible to identify a structure-sensitivity of fluorescein reduction by sodium borohydride on both nanofabricated and colloidal Au catalyst nanoparticles[22,23]. A second important trait of the nanoreactor approach that is of key interest here is the confinement of reaction product molecules formed on a single catalyst nanoparticle inside a nanofluidic channel since it prevents the rapid product dilution that is inevitable in an open reactor system, even if it is a microreactor[24,25]. Nevertheless, despite these advantages, also the nanofluidic reactor based single particle catalysis studies we have presented to date, fall short on the demand to not use a fluorescent reaction or plasmonic effects to determine single catalyst nanoparticle activity and/or chemical and structural state.

Inspired by similar challenges in the field of optical single biomolecule detection, where fluorescent labels[26] or localized surface plasmon resonance-based sensors[27] are widely used, we have recently introduced Nanofluidic Scattering Microscopy (NSM) for label-free weight and size determination of individual biomolecules freely diffusing in solution[28]. This was enabled by the intrinsically large optical scattering cross-section of nanofluidic channels and sub-wavelength interference between light scattered from a nanochannel and a biomolecule inside it. This interference significantly enhances the optical contrast of the molecule in a dark-field scattering microscopy setting and therefore makes the molecule directly visible without fluorescent labels or immobilization on a surface, as required, for example, in interferometric scattering microscopy (iSCAT)[29,30] or plasmonic single molecule sensing[27].

Here, we have taken inspiration from the NSM methods' successful application in single molecule biophysics and apply it to single particle catalysis. Specifically, we demonstrate how NSM can be used to directly image and quantify liquid concentration gradients inside up to 85 parallel nanofluidic channels, based on the corresponding refractive index contrast. Furthermore, we show how individual optically dark Pt colloidal nanoparticles trapped inside the nanofluidic channels can be visualized and counted. Finally, we demonstrate, using the example of the $H_2O_2$ decomposition reaction on single trapped Pt nanoparticles, that the local reactant concentration time evolution around a single particle can be measured *in situ* inside individual nanochannels, and how single-particle specific turn-over frequencies can be derived by quantitatively analyzing the light scattering of $O_2$ bubbles formed inside the nanochannels during reaction. While nanobubbles have been used before to investigate gas-producing reactions on nanoparticles, NSM eliminates the need for fluorescent dyes[31,32] and confines the developing bubbles to the geometry of the nanochannel, which greatly facilitates their quantitative analysis, in contrast to open surface experiments[33].

## Results and discussion

The nanofluidic chip design we employ here is the same as in our previous work and the corresponding micro- and nanofabrication was carried out using the exact same recipe we have published earlier[23]. The chips comprise U-shaped microchannels on either side of the nanofluidic system that are connected and used to transport liquid to and from that nanofluidic system (**Figure 1a**). The nanochannels are arranged in a set of 100 parallel channels, where each channel is 340 µm long and has a quadratic 150 nm x 150 nm cross-section (**Figure 1b**). To enable the trapping of colloidal nanoparticles according to the recipe we have recently established[23], each nanochannel is equipped with a 120 nm high and 1 µm long constriction ("trap") in the center. When flushed through the nanochannel and arriving at that trap, particles with a diameter larger than the 30 nm gap between the trap and the nanochannel wall are captured, while substantial liquid flow through the channel is still enabled. The entire fluidic system of the chip, including the particle trap, is etched into the 2 µm thick thermal oxide layer of a silicon wafer and sealed by the bonding of a 175 µm thick glass lid (see Methods for details). A dark-field scattering microscope image of such a chip reveals the nanochannels as bright lines due to their large light scattering cross section in the visible spectral range[28] (**Figure 1c,d**). The dark regions in the center of each channel correspond to the traps since they reduce the scattering cross section of the channels by locally reducing the channel dimensions[28].

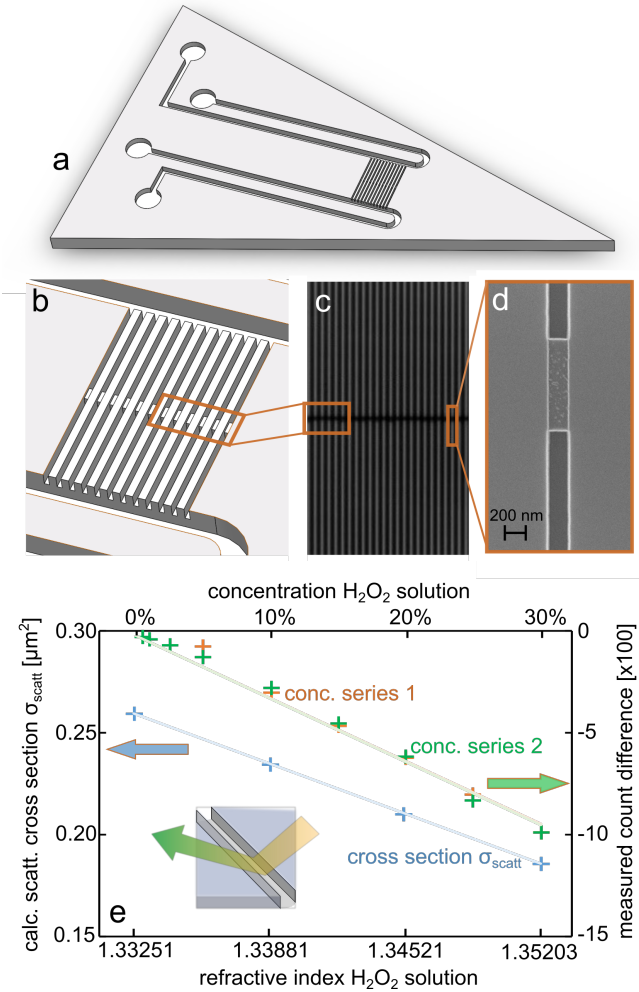

**Figure 1. Nanofluidic chip design and scattering intensity dependence on liquid refractive index.** *a) Artists' rendition of the nanofluidic chip used in the experiments. Two inlet reservoirs are connected to two microchannels that connect to either side of a set of parallel nanochannels. They enable liquid flow through the nanochannels by applying pressure to the reservoirs on one side. b) Artists' rendition of the array of parallel nanochannels that is functionalized by a constriction ("trap") in the center region to enable trapping of colloidal nanoparticles flushed through the system, provided the particles are larger than the trap in one dimension[23]. c) Dark-field optical microscopy image of an array of nanochannels with a nominal 150 nm x 150 nm cross section. The trap region that is 1 μm long and 120 nm high appears as a darker area due to the smaller scattering cross section of the nanochannel in this region[28]. d) SEM image of a single trap. e) Comparison of the theoretically calculated*

(Equation 1) dependence of the nanochannel light scattering cross section on $H_2O_2$ concentration in water inside a nanochannel with the experimentally measured dependence of nanochannel scattering intensity on $H_2O_2$ concentration. The found linear dependence is a direct consequence of the linear dependence of the refractive index of the $H_2O_2$ solution on concentration.[34]

*Concentration gradient measurements in single nanochannels*

To establish the measurement principle used in the experiments reported in this work, we first used a nanofluidic chip of the type described above with empty nanochannels, that is, without trapped nanoparticles. We flushed aqueous $H_2O_2$ solutions with concentrations up to 30% through the system and measured the difference in light scattering intensity from a single channel compared to the same channel being filled with pure $H_2O$. This revealed a linear correlation in two independent measurements in good agreement with each other (**Figure 1e**). To rationalize this experimentally identified linear dependence, we employed Mie theory and approximated the nanochannels with square cross-sections used in the experiments by an infinitely long cylinder with a diameter of 150 nm. Following the formalism described by Bohren and Huffmann,[35] and detailed in the Supplementary Information Section **SI II**, we arrive at an expression for the light scattering cross section of the nanochannel, $\sigma_{sca,u}$, upon irradiation by unpolarized light

$$\sigma_{sca,u} = \frac{A_{\emptyset}^2 k^3 L}{4} (m^2 - 1)^2 \left(\frac{1}{2} + \frac{1}{(m^2+1)^2}\right). \qquad \text{Equation 1}$$

Here, $m$ is the ratio of the refractive indices (RI) of the liquid in the channel, $n_l$, and of the $SiO_2$ medium the channel is etched into, $n_{SiO2} = 1.459$,[36] that is, $m = n_l/n_{SiO2}$, $k = 2\pi/\lambda$ is the wavenumber, $A_{\emptyset}$ the geometrical nanochannel cross section and $L$ length of the illuminated channel section. To plot and compare the theoretically calculated scattering cross section with corresponding experimental scattering intensity measurements, we assume for pure water $n_w = 1.333$,[37] and for aqueous $H_2O_2$ solutions with 10%, 20% and 30% $H_2O_2$ concentration $n_{H2O2} = 1.3394; 1.3460; 1.353$, respectively[38]. This corresponds to good first approximation to a linear dependence of the RI of water – $H_2O_2$ solutions, as reported by Hart and Ross[34]. Consequently, the value of $m$ ranges between $m_w = 0.914$ for a water-filled channel and $m_{H2O2} = 0.928$ for a channel filled with 30% $H_2O_2$ solution. This yields a linear dependence of the calculated nanochannel scattering cross section $\sigma_{sca,u}$, which is in excellent agreement with the experimentally measured linear dependence of the scattering intensity (**Figure 1e**). As the key consequence, this result means that if appropriately calibrated, NSM enables *absolute* measurements of concentration changes in liquids inside a nanochannel.

Having established this direct proportionality between the light scattering intensity from a nanochannel and the $H_2O_2$ concentration inside it both theoretically and experimentally, we will now use it to measure dynamic changes of liquid composition inside a single nanochannel in real time and at the absolute level. This is of high relevance in the specific context of this work, where we aim at investigating the formation of concentration gradients due to chemical conversion on a single nanoparticle, as well as in more general terms, to scrutinize fluid flow and diffusion in nanoconfined systems. In the first example to demonstrate this, we filled the two microfluidic systems contacting the set of nanofluidic channels on either side with pure MilliQ water and a 30% aqueous $H_2O_2$ solution, respectively, and monitored the diffusion of the $H_2O_2$ solution into five initially water filled and 340 µm long parallel nanochannels (170 µm in the field of view). A corresponding time-series of selected scattering images reveals the approaching $H_2O_2$ diffusion front as a "darkening" of the channels from the righthand side (**Figure 2a**). This is the consequence of the higher RI of the $H_2O_2$ solution compared to pure water, which means that it is closer to the RI of the nanochannel walls and thus reduces the scattering intensity of the system (*cf.* **Equation 1**). This reduction of the scattering intensity along a single nanochannel is shown in **Figure 2b** for the same time intervals as in **Figure 2a**. The establishment of a close to linear concentration profile becomes evident after 10 s. This is in good agreement with other works[39,40] that investigated the time evolution of concentration profiles between two solution reservoirs, as well as with the original diffusion laws established by Fick[41].

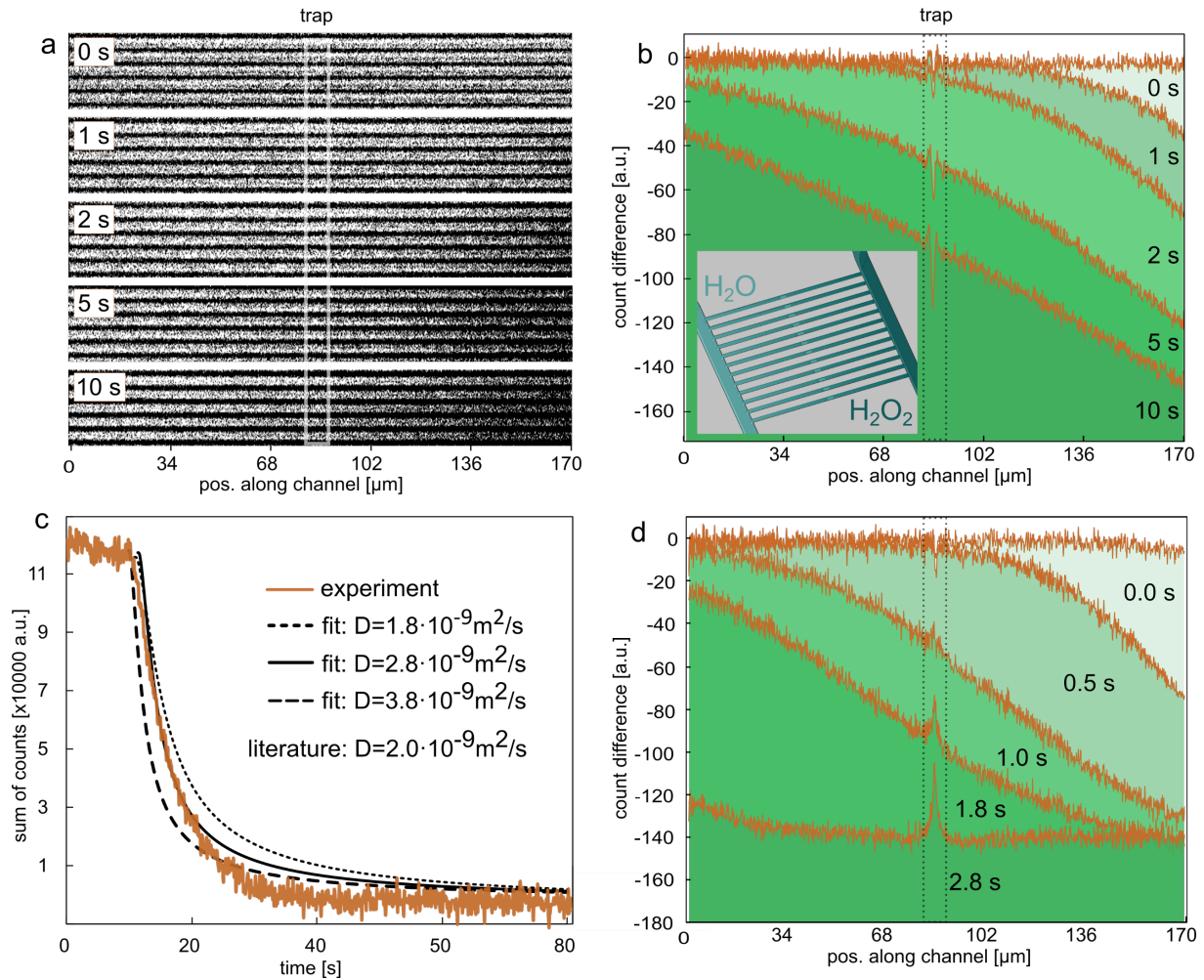

***Figure 2. Assessment of $H_2O_2$ diffusion inside nanochannels.*** *a) Dark-field scattering image time series of a set of five parallel nanochannels taken at different times after the onset of the diffusion of a 30 % $H_2O_2$ solution into the initially water filled channels. Note the approaching $H_2O_2$ diffusion front as a "darkening" of the channels from the right, as a consequence of the reduced RI difference between the channel and the liquid inside it. b) Light scattering intensity difference between a nanochannel filled with pure MilliQ water at t=0 s and the same channel upon diffusion of a 30 % $H_2O_2$ solution into it from the right. The green shaded areas under the curves indicate the corresponding integrated areas, whose time evolution is plotted in c)*

*together with corresponding time evolutions calculated using equation 2 for three different D. The best agreement between experiment and calculation is found for $D = 2.8 \cdot 10^{-9} \, m^2/s$, which is in very good agreement with the literature value[42] of $D = 2 \cdot 10^{-9} \, m^2/s$. d) Similar scenario as for b), but with convective flow due to applied pressure at the microchannel inlets.*

To further quantify the measured time evolution of the concentration profile along the nanochannel, we used it to estimate the bulk diffusion constant, *D*, of $H_2O_2$ in water (**Figure 2c**). The concentration profile can be theoretically described over time with Equation 2, which we derived from Fick's second law for a nanochannel of length *L*, when a concentration of $c_0$ is present at one end at t=0.

$$c(x,t) = c_0 \left( Erf\left(\frac{x}{2\sqrt{Dt}}\right) \Big/ Erf\left(\frac{L}{2\sqrt{Dt}}\right) \right). \qquad \qquad Equation\ 2$$

To quantitively evaluate the changing concentration profile shown in **Figure 2b**, we integrated the area under the profiles at each time point and plot the time-dependent change of this integrated area in **Figure 2c**. This reveals the rapid establishment of an almost linear profile within the first two seconds, followed by an asymptotic development towards the ideal perfectly linear profile. To extract *D* from these data, we analytically modeled the concentration profile in the channel using **Equation 2**, and applied the same evaluation scheme, that is, integration of the area under the profile curves, as done for the experimental data (cf. **Figure 2b** and **Figure S1**). Subsequently plotting the theoretically obtained curves for $D = 1.8 \cdot 10^{-9} \, m^2/s$, $D = 2.8 \cdot 10^{-9} \, m^2/s$ and $D = 3.8 \cdot 10^{-9} \, m^2/s$, we find the best agreement for $D = 2.8 \cdot 10^{-9} \, m^2/s$, which indeed is very close to the literature value[42] of $D = 2 \cdot 10^{-9} \, m^2/s$. This result is important from two perspectives: (i) it corroborates the ability of NSM to not only measure concentration changes inside nanofluidic systems but also enable the quantitative experimental determination of diffusion constants; (ii) it confirms that despite significant nanoconfinement, the macroscopic description of molecular diffusion is still valid.

In a wider perspective, it also hints at the possibility to apply NSM in experimental studies of diffusion in more extremely confined systems where molecular interactions with the nanochannel walls may become sizable, and the eventually dominant contribution, to diffusive molecular transport[43,44]. In the measurements described below, we will not solely rely on diffusion to introduce $H_2O_2$ into the nanochannels but apply a pressure of 2 bar at one of the inlets to create a convective flow. **Figure 2d** has been recorded in the same way as **Figure 2b**, but here the convective flow pushes the concentration front quickly through the channel, so that the whole length of the channel is filled with the nominal $H_2O_2$ concentration within 3 s. As a final aspect, we focus on the evolution of the scattering intensity in the trap area in **Figure 2b,d**, which is indicated by the dashed lines. We find that in this region both small positive and negative peaks evolve, which likely are the consequence of the significantly reduced scattering intensity in the trap region and the consequent negative impact on S/N in the differential image.

*Platinum nanoparticle trapping, imaging and counting*

To prepare the nanofluidic chips for measurements of the catalytic decomposition of $H_2O_2$ over single nanoparticles, we functionalized the channels by trapping citrate-stabilized spherical colloidal Pt nanoparticles comprised of small 2 – 5 nm crystallites (see **Figure 3a)** and with a mean diameter of $69.6 \pm 5.5$ nm, as evident from a size distribution histogram obtained from scanning electron microscope (SEM) images (**Figure 3b**). For the particle trapping, we filled the reservoir of the microfluidic system on the inlet side of an already water-filled chip with a diluted aqueous suspension of the Pt particles ($10^9$ particles/mL). Subsequently, we applied 2 bar pressure to the inlet side of the chip to establish a flow through the microchannels that transported the particles into the nanochannels, where they get trapped at the position of the constriction.

To monitor this process, we imaged the 84 parallel nanochannels in the field of view of the EMCCD camera at a frame rate of 10 fps and subtracted a reference image of the empty channels from each frame (**Figure 3c**). The reference image was the first image of the series of 1000 images acquired during the particle trapping. This procedure results in a time series of differential images in which the Pt particles become visible as distinct diffraction limited dark spots (**Figure 3d,e**). This is an important result because optically lossy metal nanoparticles, such as Pt or Pd, are invisible in conventional dark-field scattering microscopy in the sub 100 nm particle size range due to their localized surface plasmon resonance (LSPR) excitations predominantly decaying via absorption, rather than scattering.[45] Here, however, we propose that they become visible because of two mechanisms that likely act in concert. According to the first one, since also Pt particles scatter light, even though very little, this light may interfere with the light scattered from the nanochannel, thereby reducing the overall intensity scattered from the combined system (that is, particle plus channel) in analogy to the NSM enhancement mechanism in single molecule detection[28]. According to the second mechanism, a sizable fraction of the light scattered from the nanochannel is absorbed by the Pt particle and is thus the reason for the observed scattering intensity reduction at the position of the particle in the channel in the differential image.

As a consequence, our method offers a complement to traditional dark-field scattering microscopy of metal/plasmonic nanoparticles in the sub $\approx 50$ nm particle size range and/or for lossy metals where absorption is the dominant LSPR decay channel and thus renders them "invisible" in a traditional scattering experiment[45]. Furthermore, it also enables the tracking of the motion of such single nanoparticles inside the nanofluidic channels, as illustrated in **Figure 3e** that depicts snapshots of five channels over the course of 15 s (see also **Video SV1**). The starting point of this experiment is that particles have been trapped at the constriction by the flow applied through the nanochannels. This flow was then stopped at t = 0 s when the first image was taken. Interestingly, already after 0.1 s, the particles have started to drift away from the trap. This drift is a combination of Brownian diffusive motion and a minuscule convective flow that has its likely origin in a small difference in the static pressure induced by the liquid in the reservoirs of the microfluidic system on either side of the nanochannels. Consequently, the particles continued to exhibit a net motion in the direction away from the trap during the subsequent 15 s until we again established a distinct convective flow towards the trap that pushed the particles back towards it (**Figure 3e**).

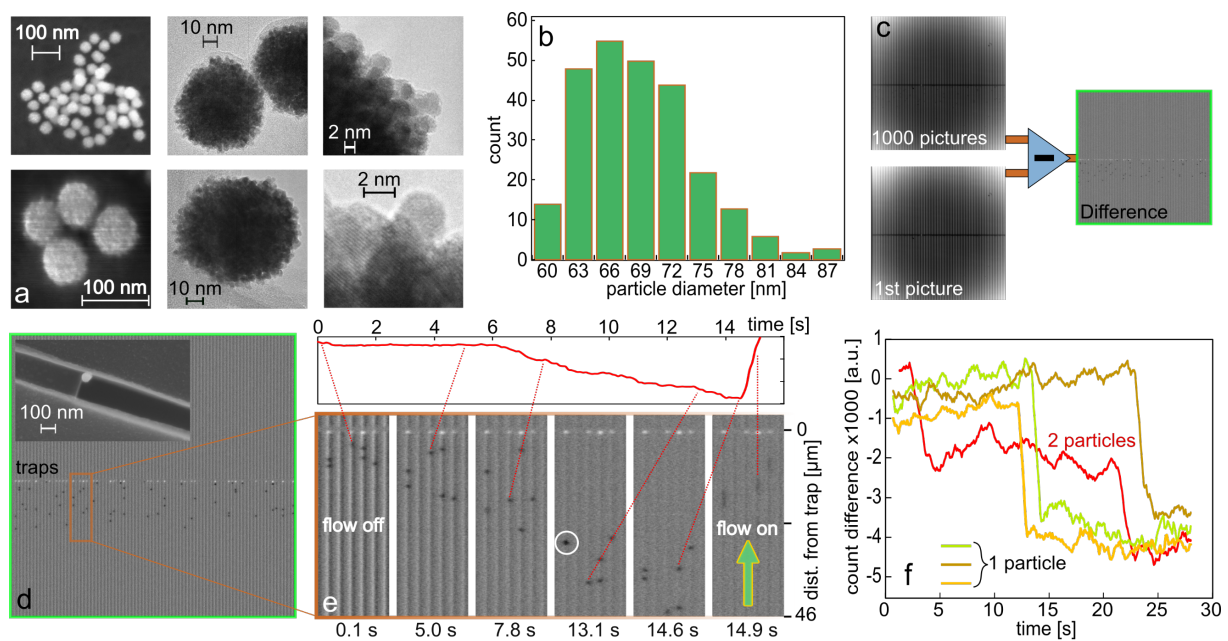

***Figure 3**. Pt nanoparticle characterization, trapping, optical detection and counting. a) SEM and high-resolution transmission electron microscopy (HRTEM) images of the citrate-stabilized Pt nanoparticles used in this work. They reveal that the particles are comprised of small 2 – 5 nm crystallites (for a collection of additional TEM images, see **Figure S2**). b) Particle diameter histogram obtained from SEM image analysis, yielding a wide size distribution with a mean particle diameter of 69.6 nm. c) Scheme depicting the generation of differential images by subtracting the first image of a measurement series, obtained when no particles had yet entered the channels, from subsequently taken images. d) Differential scattering image revealing the Pt nanoparticles as dark diffraction limited spots in the channels. The inset shows an SEM image of a 70 nm Pt particle in front of a trap. e) Time series of differential images focused on 5 parallel nanochannels that reveals the net motion of*

*nanoparticles away from the trap (bright area on top of the images) in a combination of Brownian motion and slow convective flow before reverting the flow direction in the last image at 14.9 s to push the particles back towards the trap. We also note the transient localization of two particles within a diffraction limited spot, which is reflected as a significantly darker spot in the leftmost channel at 13.1 seconds (white circle). The inset depicts the position of a single particle along the channel over time. f) Differential scattering intensity time traces of the trap region where distinct steps signal the arrival of a single Pt nanoparticle at the trap. In the red trace, the subsequent arrival of two single particles is observed. The traces shown are examples for the measurement depicted in* **Figure 5c**.

To discuss a further interesting aspect of **Figure 3e**, we focus on the leftmost nanochannel. We notice that two distinct single particles are in this channel at t = 0.1 s and that they form a "dimer" at 13.1 s where they appear as a single spot that is significantly darker compared to the images where the two particles are seen individually (white circle in **Figure 3e**). This is the consequence of the particles being transiently localized close to each other at a distance smaller than the diffraction limit of the irradiated light. Quantitatively analyzing the scattering intensity of the system at the position of the two individual particles at 5 s, where they are distinctly visible, and at the position of the particle "dimer" at 13.1 s, reveals a 40% lower scattering intensity of the dimer compared to the single particles, which hints at a $\sqrt{N}$ dependence of scattering intensity with particle number localized inside a diffraction limited spot (**Figure S3**).

As the final step, we demonstrate that this distinct reduction in scattering intensity induced by single nanoparticles inside a nanochannel enables their counting also at the position of the trap. Specifically, their arrival at the constriction is manifested as a distinct step in the time trace of the scattering intensity measured at the trap. This enables the detection of the arrival of single and multiple individual nanoparticles inside a single nanochannel in analogy to our previous work,[23] however, here also for optically dark particles (**Figure 3f**). This is important to ensure that the desired number of particles inside each nanochannel can be verified prior to the catalysis experiments we discuss below, where we target a situation with as many channels as possible being functionalized with a single nanoparticle only.

*Catalytic $H_2O_2$ decomposition over single Pt nanoparticles*

Having established the trapping and imaging of the single Pt nanoparticles, as well as the measurement of concentration changes in a solution inside a single nanochannel, we now apply the developed system to investigate a catalytic reaction. For this purpose, we have chosen the catalytic decomposition of $H_2O_2$ on Pt, which takes place according to a two-step cyclic mechanism[46]. In the first and rate limiting step a $H_2O_2$ molecule reacts with the Pt surface and forms a chemisorbed oxygen on the Pt surface, Pt(O), and a water molecule, $H_2O$, that desorbs from the surface. In the second step, a second $H_2O_2$ molecule reduces the Pt(O) back to metallic Pt by forming $O_2$ and $H_2O$, which both desorb from the catalyst and thereby close the cycle. This yields an overall reaction that can be written as

$$2H_2O_2 \xrightarrow{Pt} 2H_2O + O_2.$$   *Equation 2*

Focusing on the Pt nanoparticles we use in this work, due to their structure that features relatively small crystallites, they are characterized by a large surface area that features both edges and terraces at relatively high abundance. Hence, they are expected to be highly active due to the interplay between high (terraces) and low (edges) coordination sites, which, in combination, keep binding energies of reactants and intermediates at a moderate level and thus sees to that the activation energy of the rate determining step and the reactant surface coverage is relatively low[46]. Furthermore, the highly structured surface facilitates effective detachment of oxygen bubbles from the particles[47]. Such bubbles are expected to form on the surface during reaction if the reaction rate is high enough to produce more $O_2$ than can be dissolved in water.[47]

Projecting the potential of the $H_2O_2$ decomposition reaction to form $O_2$ bubbles onto our nanofluidic reactor system at hand (**Figure 4a**), after the detachment of small $O_2$ bubbles from the surface of a trapped Pt nanoparticle inside a nanochannel, there are two possible scenarios according to which they can condense into a larger bubble when the $O_2$ solubility limit (1.22 mol/m$^3$ in water[42]) is reached. *Scenario I:* following the direction of the convective flow through the channel, a bubble is formed downstream of the particle and across the trap constriction (**Figure 4b**). *Scenario II:* growing against the direction of the convective flow through the channel, a bubble is formed upstream of the constriction, either between the trap and the particle or upstream of the particle (**Figure 4c**).

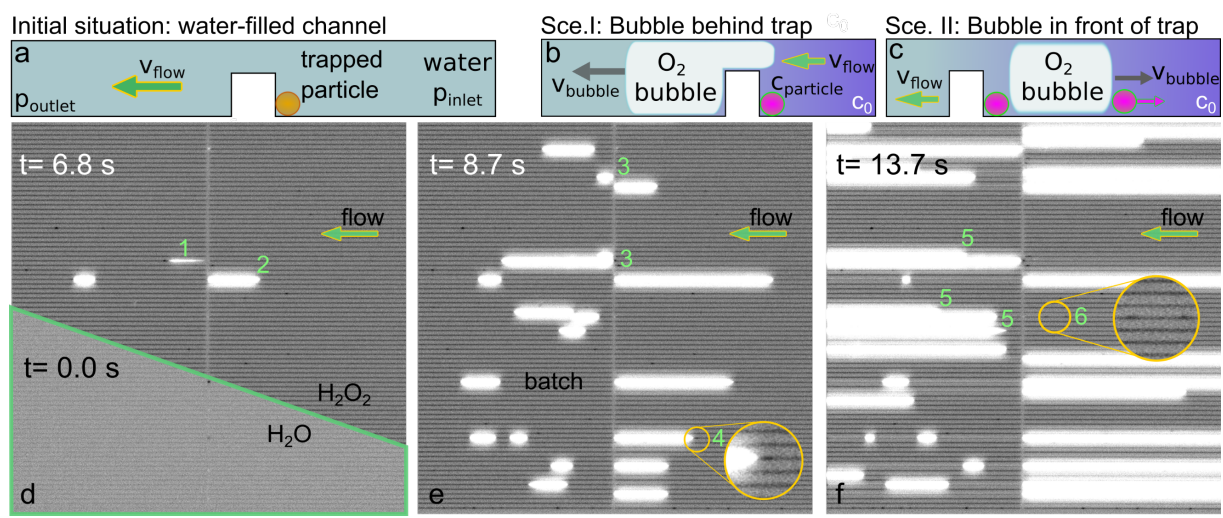

***Figure 4. $H_2O_2$ decomposition reaction on Pt nanoparticles and $O_2$ bubble formation.*** *A) Schematic of the initial situation with a trapped Pt particle and the entire nanochannel filled with water. B) Scenario I: upon convective inflow of $H_2O_2$ into the nanochannel from the righthand side an $O_2$ gas bubble forms downstream of the particle due to the catalytic decomposition of $H_2O_2$ on the particle surface. C) Scenario II: the $O_2$ bubble either forms between the trap and the particle or upstream of the particle. In both cases it grows upstream against the convective inflow of reactant. D) Darkfield scattering images of the completely water-filled nanochannel system after Pt nanoparticle trapping (green frame) and the system*

*after complete liquid exchange to 30 % $H_2O_2$ manifested by the darker appearance of the channels at t=6.8 s. We also note the onset of $O_2$ bubble formation on two particles (highlighted by label 1 and 2 – see main text for details) manifested as areas of intense light scattering. E) Darkfield scattering image of the nanochannel system at t=8.7 s when multiple additional bubbles have formed. The labels highlight specific scenarios discussed in detail in the main text. F) Darkfield scattering image of the nanochannel system at t=13.7 s. The labels highlight specific scenarios discussed in detail in the main text. For all images we note that dark spots not positioned close to the traps are artefacts caused by structural defects of the nanochannels generated during their fabrication (that is, they are not Pt particles). The field of view corresponds to 170 μm x 170 μm.*

To investigate these two possible scenarios, we trapped Pt nanoparticles in the nanochannels and monitored the trapping process using the concept outlined above (cf. **Figure 3f**). Subsequently, we exposed the trapped particles to a flow of 30 % $H_2O_2$ in the direction towards the trap. This resulted in 14 parallel nanochannels exhibiting Scenario I, whereof one channel was occupied by a single Pt nanoparticle, ten channels were occupied by 2 nanoparticles and three channels were hosting 3 or more particles (**Figure S4**). Furthermore, twelve nanochannels exhibited Scenario II. Finally, some particles also attached to the nanochannel wall before reaching the trap (see **Figure S5** for an example).

After the functionalization of the chip with the Pt particles, we investigated the bubble evolution via a series of dark-field scattering images taken from the parallel nanochannels (**Figure 4d-f**). For this experiment, with the Pt particles at the trap, we started out with water in one microfluidic channel and in the nanochannels, and with 30 % $H_2O_2$ in water in the second microchannel. Subsequently, we established convective flow from the $H_2O_2$ side through the nanofluidic system by applying 2 bar pressure on the $H_2O_2$ inlet reservoirs. At start, the particles were still fully immersed in water, as one can see from the overall relatively bright image which turns distinctly darker as all channels are filled with 30 % $H_2O_2$ (**Figure 4d**). Subsequently, when $H_2O_2$ had been flushed in, $O_2$ bubbles started to form as the $H_2O_2$ decomposition reaction was initiated. In general, the appearance of an $O_2$ bubble is manifested as a dramatic increase in scattering intensity since it locally expels the liquid from the nanochannel and thereby changes the refractive index contrast of the system significantly. From here forward, we will discuss in detail several particularly interesting events specifically labeled in green in **Figure 4d-f**.

Label 1 highlights an event occurring when the reaction starts on the first particles in the system, where an initial $O_2$ bubble is formed at a Pt particle caught at the trap, but then detaches from that particle and moves through the nanochannel driven by the flow (**Figure 4d**). Label 2 highlights the formation of an $O_2$ bubble according to Scenario II in a neighboring channel. After 8.7 s more bubbles have started to form at an increasing number of Pt particles (**Figure 4d**), and we can see examples of Scenario I and II (Label 3 in **Figure 4e**). Interestingly and marked by label 4, a Pt particle seen as a dark spot in the zoom-in, is being pushed out by the bubble that in this case grows between the particle and the trap. **Figure 4f** shows the later stages of the measurement series, with several bubbles having reached the end of the field of view and with many channels exhibiting Scenarios I and II. Label 5 marks large bubbles in Scenario I that have detached from the trap and are on the way to being flushed out of their channel. Label 6 highlights two particles (dark spots in the zoom-in) that have diffused away from the trap since the convective flow through the channel is significantly reduced by the $O_2$ bubble downstream in the same channel. An interesting variant of Scenario I is shown in **Video SV2**, where in one channel bubbles form in the direction of flow but then detach rather quickly from the trap region, such that several bubbles form in rapid succession from a single particle, before a Scenario II bubble is established.

To further analyze the observations made, it is interesting to discuss them from the perspective of the two scenarios depicted in **Figure 4b,c**. In Scenario I (**Figure 4b**), where the $O_2$ bubble grows in the direction of the applied $H_2O_2$ flow and downstream of the particle trap, the situation is rather straightforward since $H_2O_2$ has unrestricted access to the particle via convective flow and diffusion, and since the particle remains in position at the trap during the entire experiment while the bubble extends on the other side of the trap towards the outlet microchannel. However, we also note that the bubble growing downstream of the trap will reduce the convective flow through the channel gradually as it grows, since it increases the flow resistance of the system (**Figure S6**). At the same time, it will not completely block the flow through the channel due to the hydrophilicity of the channel inner walls, which ensures that a thin layer of liquid is sustained between the $O_2$ bubble and the nanochannel wall.

Scenario II (**Figure 4c**) is more complicated because: (i) if the bubble nucleates between the trap and the particle the bubble may expel the particle from the channel during growth by pushing it ahead (cf. label 4 in **Figure 4e**); (ii) if the bubble nucleates and grows away from the particle against the convective flow, it effectively blocks a large fraction of the $H_2O_2$ inflow by occupying a large fraction of the channel cross-section. However, since it does not block the inflow completely due to the liquid layer between wall and bubble mentioned above, a continuous, yet reduced, supply of $H_2O_2$ ensures the continued $O_2$ formation and bubble growth that we observe in the experiment, together with the $H_2O_2$ that had been flushed past the particle before bubble development. However, the rate of supply might decrease as the bubble expands due to the correspondingly increasing flow resistance (**Figure S6**).

Finally, it is also interesting to estimate to what extent the $O_2$ bubbles fill out the channel cross section and thereby get an indication for the dimensions of the liquid layer between the bubble and the channel wall. To do this, we compared the light scattering intensities of completely air-filled channels with channels filled with an $O_2$ bubble formed by the $H_2O_2$ decomposition reaction. This analysis yields a filling factor of 75.5 % of the total nanochannel cross section by the $O_2$ bubbles (**Figure S7**), which translates into an estimated liquid layer thickness of 9 nm between the nanochannel wall and a bubble inside it.

We also note that analysis of the bubble development directed against the flow (Scenario II) provides comparable results to Scenario I (**Figure S8**). However, for the further quantitative study of the catalytic reaction presented below, we chose to exclusively focus on Scenario I, where the bubble growth occurs on the downstream side of the particle and the trap, and the tracing the spatial extension of an $O_2$ bubble along the nanochannel over time provides insight into the amount of $O_2$ produced in the $H_2O_2$ decomposition reaction on a single Pt nanoparticle surface. In fact, since the filling factor of the bubble and the exact geometry of the nanochannel are known, the absolute amount of gas phase $O_2$ produced over time can be accurately calculated. Consequently, we can define a bubble extension speed (BES), which is the change of bubble length in pixels per second, multiplied with the scale of 0.166 µm per pixel. This BES is then directly proportional to the current reaction rate of the particle. The influence of the solubility of the formed $O_2$ in water is shortly discussed in SI **Section II**.

Even though the measurement shown in **Figure 4** provided many $O_2$ bubbles for evaluation (**Figure S4**), only one channel was functionalized with only a single nanoparticle. We therefore repeated the same experiment, including the particle trapping, in a fresh identical chip. This resulted in 9 channels being functionalized with a single Pt nanoparticle and one channel with two Pt particles (**Figure S7**). The remaining channels were either empty or accumulated larger particle numbers. Subsequently flushing the nanochannel system of the chip with an $H_2O_2$ solution induced $O_2$ bubble formation due to the catalytic decomposition of $H_2O_2$ on the trapped Pt particles. An example of an accordingly obtained single Pt nanoparticle BES time trace is shown in **Figure 5a**, together with the scattering intensity integrated along 85 µm of the channel up- and downstream of the trap, respectively. A BES trace is terminated when it cannot be measured in a consistent way anymore. This is the case when the bubble extends beyond the edge of the field of view of the microscope. Furthermore, we also consider the detachment of a bubble from the trap as an end of the BES trace since in this case the connection to the particle is lost and the bubble is moving through the channel (see also **Figure S9**).

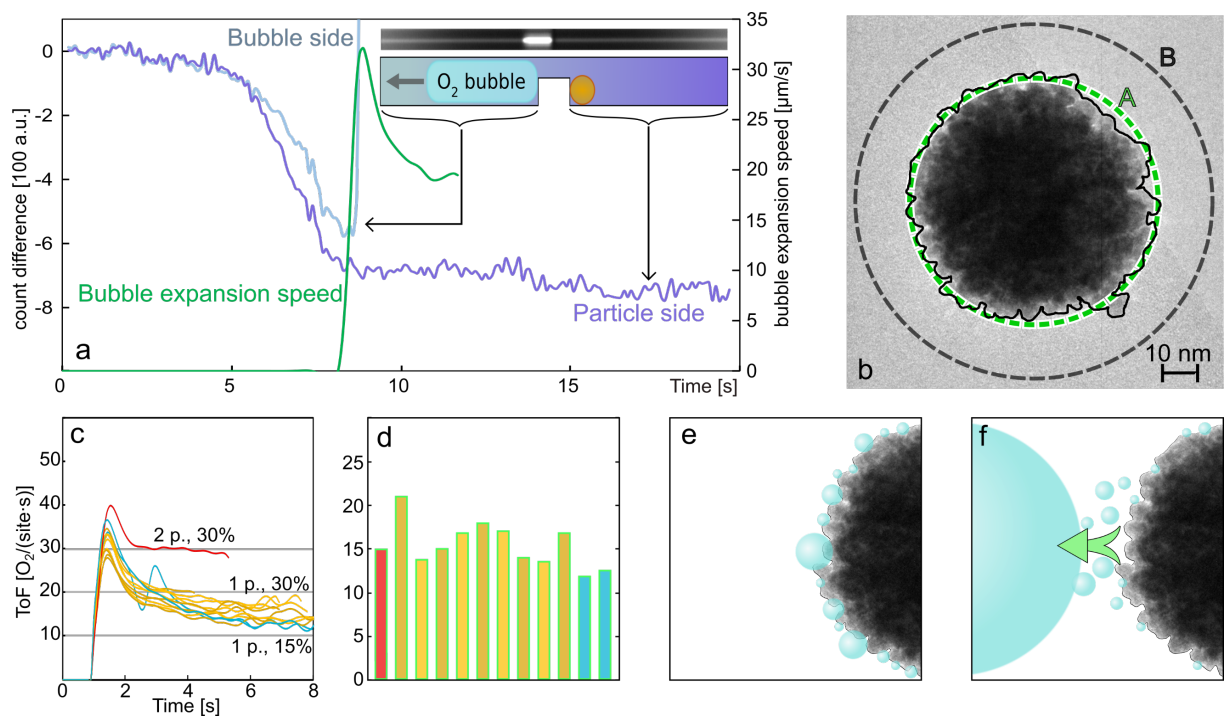

***Figure 5****. Single particle reaction turnover frequencies (ToF) and local reactant concentrations in the nanochannel. a) Change of nanochannel scattering intensity over time with respect to the scattering intensity of the fully water filled channel. The signal was integrated along 85 μm of a channel up- (purple) and downstream (grey) of the trap (see inset) and is plotted together with the bubble expansion speed (BES, green) for a bubble forming downstream of a single Pt nanoparticle trapped at the constriction (see second inset for corresponding dark field scattering image snapshot). A video of this process is included in the SI (**SV3**), together with a figure that shows characteristic points in time (**Figure S10**). The time traces start at t = 0 s when the channel is entirely filled with water and subsequently flushed with 30 % H₂O₂. Note the initial reduction of the scattering intensity signals prior to the onset and steep increase of the BES, as well as the ca. 1 s time delay in the scattering intensity decrease between the up- and downstream signals that is the consequence of the convective transport of H₂O₂ through the channel to replace the water (see also see **Figure S11**). b) HRTEM image of a single Pt nanoparticle representing equivalent particles trapped in the nanochannel. It is characterized by a rough surface dictated by the 2-5 nm sized crystallites the particle is comprised of. The dashed green line (A) depicts the particle circumference, whereas the dashed black line (B) depicts the circular equivalent of the rough particle surface*

*outline (thin black line). The corresponding 3D surface areas are A =15394 nm² and B= 31416 nm². c) ToF time traces derived from BES determined from individual nanochannels and using a single particle surface area B. Nine channels are decorated with a single Pt nanoparticle and one nanochannel is decorated with 2 Pt particles, resulting in an almost doubled ToF. All particles were measured simultaneously in the same experiment using 30 % $H_2O_2$ concentration. Also shown is the ToF time trace for a single Pt nanoparticle obtained in 15 % $H_2O_2$ in a separate experiment. The rise of all ToF traces was shifted to 1s on the time axis to enable a comparison of their development. All analyzed particles exhibited bubble formation according to Scenario I. d) Bar plot of the steady state ToF extracted from panel c) scaled to the level of a single particle in the case of two particles in the channel. e) Schematic depiction of the initial formation of small $O_2$ bubbles on the Pt nanoparticle surface. f) When a critical size and surface coverage of the small $O_2$ bubbles is reached, they coalesce in an avalanche-like way into a large bubble that subsequently detaches from the particle. This is the reason for the initially observed very rapid bubble development in the nanochannels that after a peak in BES reaches a lower steady state where the BES is solely determined by the continuous catalytic production of $O_2$ on the particle surface.*

It is now interesting to discuss in detail the time evolution of the BES and the difference in the integrated channel scattering intensities extracted both up- and downstream of the trap with the Pt particle (**Figure 5a**). Focusing first on the scattering intensity difference traces, from t = 0 s they initially remain constant up- and downstream of the trap since the channel in this state is entirely filled with water and no $O_2$ bubble is present. Subsequently, after ca. 5 s, the scattering intensity starts to decrease and signals the arrival of the 30% $H_2O_2$ solution in the respective channel sections. Interestingly, we can resolve a small delay of 1 s (**Figure S11**) between the up- and downstream side of the trap, in agreement with the convective flow of the $H_2O_2$ through the nanochannel. Furthermore, focusing on the BES trace, we notice that its onset nicely coincides with the 30% $H_2O_2$ concentration being fully established in the particle side of the channel. This indicates that the bubble nucleates and grows once the catalyst particle is fully immersed in the reactant solution. Interestingly, however, it is not the case for all studied particles that bubble formation is initiated first when the full 30% $H_2O_2$ concentration is established at the particle, since for several of them bubble formation sets in already earlier. We interpret this as that highly reactive single, or small agglomerates of several nanoparticles can produce enough oxygen to nucleate an $O_2$ bubble already at lower $H_2O_2$ concentrations, while particles with a lower activity may need more time to decompose sufficient $H_2O_2$ for the formation of a visible bubble in the nanochannel. In line with this reasoning, we therefore also observe that the single particles that exhibit an early onset of bubble formation often are those who subsequently exhibit the highest BES, once the full 30% $H_2O_2$ concentration is reached in the channel, which indicates significant differences in (apparent) activity between individual nanoparticles (**Figure S4b, Figure S8b**).

As a second aspect, it is interesting to analyze how the $H_2O_2$ concentration in the nanochannel evolves while the decomposition reaction is running on the particle because it sheds light on whether concentration gradients are being formed due to reactant conversion and if the catalyst is operated in a mass-transport or kinetically controlled regime. Accordingly, monitoring the nanochannel scattering intensity time trace upstream of the particle, that is, in the direction of reactant supply, reveals that it remains constant at the same level as at the onset of bubble formation (**Figure 5a**). Since we, based on the earlier calibration of the scattering intensity towards $H_2O_2$ concentration (cf. **Figure 1**), know that the measured signal corresponds to the nominal 30% $H_2O_2$ in water in the inlet, we can conclude that the concentration upstream of the particle remains constant and is not depleted over time. This is an important result, because it is the experimental proof that the combination of convective flow and rapid $H_2O_2$ diffusion towards the catalyst nanoparticle is fast enough to prevent the formation of a gradient despite the tiny volume of the nanochannel. Hence, it confirms that we are operating the catalyst particle in the kinetically limited regime and at a well-defined concentration. This, in turn, means that the particle-specific reaction rates we determine and discuss below are a direct consequence of single particle structure and activity. Further proof for our assumption of kinetic reaction limitation can be found in Scenario II cases. Here, the particle is moving towards the $H_2O_2$ supply (upstream in the nanochannel) while the resulting BES and ToF traces (see **Figure S8b**, also **Video SV4**) show no significant difference when they are compared to Scenario I or Scenario II data where the particles remain at the trap. As additional key points, we note that once the bubble has formed, the scattering intensity on the downstream side raises rapidly, meaning that we no longer can trace the $H_2O_2$ concentration. Secondly, during the initial phase of bubble formation and growth, we observe a distinct maximum in the BES before it first slows down significantly and rapidly, and then converges towards a steady state. We will discuss the origin of the distinct peak in BES further below. Here, we already note that the slight continuous decrease of the BES in the quasi-steady state regime (after 4 s in **Figure 5c**) may have its cause in a slow overall decrease in the $H_2O_2$ concentration in the chip due to reaction conversion.

As the next step of our analysis, we now attempt to convert the measured single nanoparticle BES into a turnover frequency (ToF) per site and second. To do that, beyond knowing the amount of $O_2$ formed per unit time, we need to determine the surface area of the Pt particles at hand to derive an estimate of the number of active sites. We resort to a high-resolution TEM image of a single Pt nanoparticle, which reveals that the surface is rough due to the particle being comprised of small crystallites (**Figure 5b**). This in turn means that the real surface area is much larger than the one corresponding to a smooth sphere with 70 nm diameter (green dashed line in **Figure 5b**). To estimate a more realistic surface area, we thus draw the outline of the 2D projection of the rough particle surface seen in the TEM image (solid black line in **Figure 5b**) and convert it into a smooth circle with corresponding circumference (dashed black line in **Figure 5b**). This analysis reveals that the true surface area of our particles with nominal

≈70 nm diameter (surface area 15394 nm$^2$ for a smooth sphere) can be approximated by a smooth spherical particle with ≈100 nm diameter and 31416 nm$^2$ surface area.

At this point, it is necessary to discuss the reaction mechanism of $H_2O_2$ decomposition and the possible influence of the surface structure, since there are multiple pathways that lead to the production of $O_2$ and water[46]. As reported by Serra-Maia et. al.[46], the rate limiting step is the dissociation of an $H_2O_2$ molecule into a surface-bound oxygen and $H_2O$. This happens preferentially on the highly coordinated (111) and (100) terrace sites because the corresponding binding energies for oxygen are significantly higher than on edge/corner sites, which thereby lowers the activation barrier of the rate limiting step on the terraces. Given that the particles in our experiments are composed of crystallites in the few nanometers range, as a rough estimate, we assume that half of their surface is comprised of terrace sites and that the other half corresponds to edge/corners. For our estimation of the ToF below, we therefore assume that 50% of the total number of surface atoms are active sites for the $H_2O_2$ decomposition reaction[48] (see also **Figure S12**)

Based on the above assumptions, we now use the experimentally determined BES (in m/s) and the geometric cross section area of the nanochannel, $A$ = 150 nm x 150 nm, to arrive at the following expression for the ToF per site per second of a Pt nanoparticle inside a nanochannel (a more detailed derivation is given in **SI Section II**)

$$ToF = \frac{0.755 * A * BES * N_A}{V_{O2}^{mol} N} \qquad \text{\textit{Equation 3}}$$

where, $V_{O2}^{mol}$ = 22.39 mol/l is the molar volume of $O_2$ at atmospheric pressure, $N_A$ is Avogadro's constant, $N$ is the estimated number of active sites for the nanoparticle of interest. To determine $N$, we use the particle analyzed in **Figure 5b**, for which we calculated at the lower end a surface area for a sphere of 70 nm diameter to 15394 nm$^2$ and for a sphere of 100 nm diameter at the high end a surface area of 31416 nm$^2$. Assuming an atomic surface density of $1.53 \cdot 10^{19}$ atoms per $m^2$ for Pt, calculated from the interatomic distance[49], this results in a total of 235525 surface atoms for the lower and 480664 surface atoms for the upper surface area limit, respectively. Finally, applying the estimation that only 50 % of these atoms are located in the terrace sites that are important for the rate-limiting step of oxygen adsorption[46], we arrive at an estimated range of 117763 to 240332 active sites on our particles. We also note that the 0.775 prefactor has its origin in the 75.5% filling factor for the bubble in the channel that we estimated in **Figure S7**.

By applying these numbers and using the upper particle surface area limit (surface B in **Figure 5b**), we can now calculate ToFs for the 9 single Pt nanoparticles trapped in individual nanochannels and plot their ToF time evolution for a 30% $H_2O_2$ concentration in the reactant solution together with two single particles measured in 15% $H_2O_2$ (separate experiment) and with two particles trapped in a single channel measured in 30% $H_2O_2$ (**Figure 5c**).

All these ToF time traces have in common that they exhibit an initial rapid rise of the reaction rate (which is the consequence of the correspondingly observed rapid rise of the BES mentioned above, cf. **Figure 5a**), which subsequently drops and converges to a reasonably constant value after approximately 5 s. Focusing on this relatively stable regime first, we can extract the corresponding ToF values for each single particle (**Figure 5d**) and the mean value for each condition. For the single particles in 30% $H_2O_2$ we find ToF = $16 \pm 2.3$ site$^{-1} \cdot$ s$^{-1}$, for the single particles in 15% $H_2O_2$ we find ToF = $12 \pm 0.3$ site$^{-1} \cdot$ s$^{-1}$ and for the two particles in 30% $H_2O_2$ we find ToF = $14.7$ site$^{-1} \cdot$ s$^{-1}$. In combination with the fact that we know from the channel scattering intensity measurements that no concentration gradients are formed in the channel during reaction, and we thus do not expect mass transport limitations, the values imply that the lower ToF found for 15% $H_2O_2$ is the consequence of a concentration dependent $H_2O_2$ coverage on the particle surface as the reason for the reduced reaction rate at lower concentration, since it has been found that the $H_2O_2$ decomposition reaction is first order in terms of $H_2O_2$ concentration[50]. It is also understandable that we see a higher BES when two particles are in one channel, but when we consider the ToF for this scenario, it is slightly lower than the mean single particle value but still within the standard deviation. This is possibly due to obstruction of surface sites by the neighboring particle and the positioning of the two particles in front of the trap. As a further observation, we note the significant spread in ToF between individual nanoparticles (**Figure 5c,d, Figure S4c, Figure S8b**). This can primarily be attributed to their individual structure, mostly in terms of size, since the particles exhibit significant size distribution (cf. **Figure 3a,b** and **Figure S2**), while at the same time being structurally stable upon exposure to reaction conditions(**Figure S2b,c**).

The fluctuations in the ToF traces visible in **Figure 5c** can be partly attributed to two main effects. At the macroscopic scale, it is to a combination of slight instabilities of the experimental setup due to vibrations and (transient) defocusing, and the fact that the BES used to calculate the ToF is determined by counting how many pixels of the channel after the trap exceed a certain level of brightness. This means that slight motion or transient defocusing of the image leads to a transient change in the number and/or brightness of the pixels counted as part of the bubble, and thereby to a transient in/decrease of the BES, and consequently of the ToF. At the microscopic scale, there is a second reason for apparent ToF fluctuations, which originates from bubble movement along the channel that is not completely monotonous. An extreme case can be seen in **Figure 5c**, where between 2 s and 4 s the ToF of a particle in 15% $H_2O_2$ exhibits first a rapid decline that is followed by sudden increase before settling again at a normal ToF trace. The reason is that the growing bubble gets transiently stuck at a defect in the channel wall (decreasing therefore transiently the ToF) before it comes loose (sudden increase in ToF) to then continue as expected (see **Video SV5**, lower channel).

As the final aspect, it is relevant to discuss the observed distinct peak in BES and derived ToF after onset of the reaction (**Figure 5c**). We propose that this peak has its origin in the initial phase of small bubble nucleation, growth and detachment on/from the nanoparticle surface. Specifically, as the reaction is initiated and $O_2$ starts to be produced, many small $O_2$ bubbles nucleate on the particle surface[47] and start to grow (**Figure 5e**). Once they reach a critical size, and stimulated by the rough surface of our nanoparticles, these bubbles start detaching in an avalanching way from the surface[47] to subsequently coalesce into a single large bubble in solution that grows in the nanochannel where it is monitored in our experiment (**Figure 5f**). In this initial stage of formation and growth of this large bubble in the channel, its growth rate is not directly determined by the $O_2$ production rate of the catalyst surface, but instead dictated by the number and volume of available small $O_2$ bubbles that the catalyst already has produced. In other words, in this phase, the BES reflects the rate of the initial coalescence of small $O_2$ bubbles into a large one, rather than the $O_2$ formation rate on the surface. Since this rate is determined by the amount of $O_2$ already produced, it transiently exceeds the $O_2$ formation rate of the catalyst until a steady state is reached, where all processes are in equilibrium and thus the $O_2$ formation rate on the particle surface is proportional to BES.

As a next step, we attempt to compare the obtained single nanoparticle ToF values to the literature. This, however, proves difficult since a wide range of both catalyst materials and reaction conditions are reported. Selecting the most relevant ones, Liu et. al.[51] reported the amount of $O_2$ produced by silica-based nanosheets with sub-2 nm Pt nanoparticles in 3 % $H_2O_2$ in water and found an $O_2$ production rate of 343 ml/(min $\bullet$ g) by using the weight of the used catalyst as reference. Correspondingly calculating the volume of $O_2$ produced by a single nanoparticle (weight $3.85 \bullet 10^{-15}$ g) in our experiment, we find a rate of 1153 ml/(min $\bullet$ g). While different by a factor of 3, this result rationalizes our single particle ToFs in a reasonable way since the Pt loading in Liu et al. was only 0.4 wt. %. In another example, Laursen et al.[52] studied $H_2O_2$ decomposition on a Pt foil for which they report an $O_2$ production rate of $3.2 \bullet 10^{-2}$ mol/(s $\bullet$ m$^2$) for a 1% $H_2O_2$ solution at high pH. This is to be compared to the mean ToF for a single particle (16 site$^{-1}$ • s$^{-1}$) that we obtained at 30 times higher $H_2O_2$ concentration and at lower pH that translates into and $O_2$ production rate of 2 • 10$^{-4}$ mol/(s • m$^2$). Finally, Serra-Maia et. al.[46] investigated Pt nanoparticles ranging from 22 nm to 3 nm in size, but only at very low $H_2O_2$ concentrations of 0.001 mol/l$_2$, which is almost four orders lower than our 30 % $H_2O_2$ that translates into 9.7 mol/l. However, in their various papers[50,53,54] they report $H_2O_2$ decomposition rates between 10$^{-6.5}$ mol/(s • m$^2$) and 10$^{-1.5}$ mol/(s • m$^2$), depending on catalyst treatment and reaction condition, which puts our value of 2 • 10$^{-4}$ mol/(s • m$^2$) well within their range, and thus corroborates our value as very reasonable.

*Spontaneous batch reactor formation*

On rare occasions, we have observed a scenario where an $O_2$ bubble that has formed according to Scenario I detached from the trap and was swept downstream by the convective flow until it got stuck at a defect in the nanochannel wall. Simultaneously, a second $O_2$ bubble formed on the Pt particle still localized at the trap, but this time according to Scenario II, where the bubble grows upstream of the particle against the flow (**Figure 6a**, **Video SV6**). This creates an interesting situation where the two bubbles effectively enclose a segment of the channel with the catalyst trapped inside. Since the $H_2O_2$ enclosed in this segment is on one hand rapidly consumed by the reaction on the catalyst and on the other hand only slowly resupplied via the narrow layer at the bubble-channel wall interface, this scenario creates a "batch reactor". Using the nanochannel scattering intensity signal measured from such a batch reactor segment, we can therefore *directly* measure the consumption of $H_2O_2$ by the catalyst over time, while simultaneously assessing the BES of the bubble that grows upstream. **Figure 6b** displays corresponding experimental time traces measured for a batch reactor that featured several citrate-covered Pt nanoparticles in the trap and occurred during the same experiment as discussed above (cf. **Figure 4e**). Specifically, the figure shows the time evolution of the $O_2$ production rate derived from the BES together with the $H_2O_2$ concentration evolution downstream of the trap and during the initial phase of the experiment upstream from it. The concentration traces exhibit the expected course of rapid initial increase when the $H_2O_2$ is flushed into the channel. Also as before, the onset of the upstream bubble growth indicated by a measurable BES, occurs when the $H_2O_2$ reaches its maximum 30 % value and shows the typical initial peak (cf. **Figure 5**). This initial peak in the bubble evolution is however not represented in the concentration trace from the "batch" area downstream of the particle, (see **Figure 6b** at 63 s), thereby corroborating our hypothesis that the initial bubble formation is a particle-surface-controlled process and not directly dependent on concentration fluctuations (cf. **Figure 5e,f**).

As the first key difference compared to the experiments above with the "flow reactors", we notice a transient disturbance of the $H_2O_2$ concentration curve over the course of 5 frames that signals the rapid formation and detachment of the downstream bubble, which is terminated when the upstream bubble starts to form. Accordingly, the time trace of the $H_2O_2$ concentration measured upstream of the bubble is terminated at the onset of the upstream bubble formation (since it cannot be measured anymore due to the brightness of the bubble), which nicely coincides with the detachment of the downstream bubble. As the key observation from here forward, and in distinct contrast to the experiments discussed above, the $H_2O_2$ concentration we measure *inside* the batch reactor does *not* remain constant at the set level of 30 % but reduces with time to reach almost zero after ca. 18 seconds of reaction. At the same time, the rate of $O_2$ evolution keeps dropping as the concentration decreases. This is the consequence (i) of the catalyst particles rapidly consuming almost all available $H_2O_2$ inside the batch reactor and (ii) of the limited inflow of fresh reactant along the upstream bubble – channel wall interface not being sufficient to resupply the $H_2O_2$ consumed by the catalyst. We note here that the small resupply of $H_2O_2$ can occur via convective flow from the upstream side of the channel past the developing bubble as pressure is still applied to the nanochannels, but also via diffusion from the up and downstream side, as the whole nanochannel system has been filled with the $H_2O_2$ solution before bubbles started to form (visible as overall decrease in brightness in **Figure 4d**).

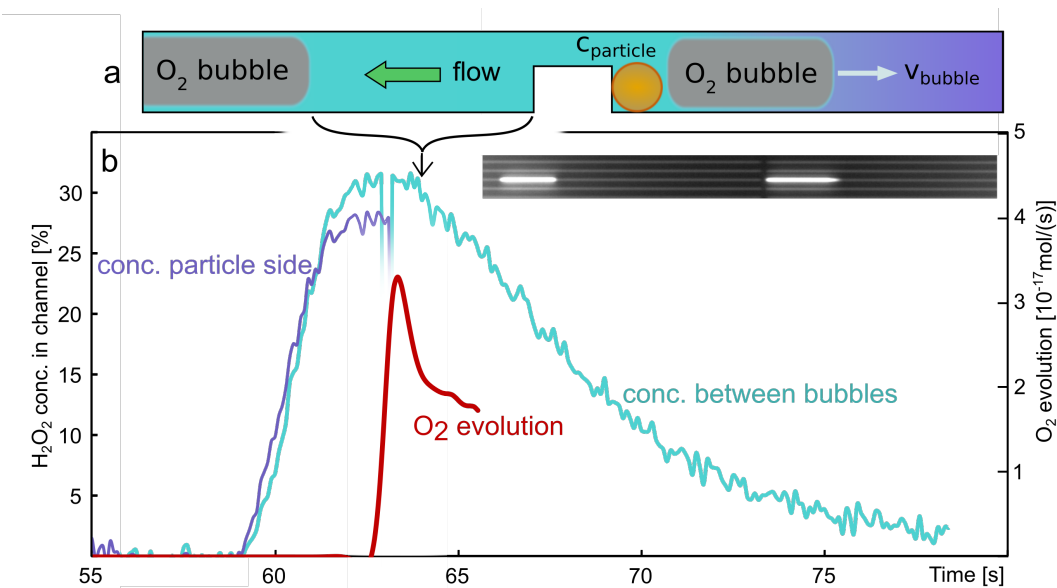

*Figure 6. **Spontaneous batch reactor formation.** a) Schematic depiction of a spontaneous batch reactor forming in a nanochannel, when a particle is trapped between two $O_2$ bubbles that to a considerable extent block inflow of reactants. b) Time evolution of the $H_2O_2$ concentration derived from the scattering intensity change measured between the bubbles of a spontaneously formed batch reactor (teal) and upstream of the particle before the onset of upstream bubble formation (purple). These time traces are plotted together with the $O_2$ evolution rate (red) derived from the BES of the upstream bubble. Note the transient disturbance of the $H_2O_2$ concentration curve at ca. t = 63 s over the course of 5 frames that signals the rapid formation and detachment of the downstream bubble. The inset shows a dark-field image of the bubbles enclosing the batch reactor (see also **Video SV6**).*

This formation and observation of a nearly closed off batch reactor within a nanochannel is an important result because it (i) is a direct and label free measurement of reactant conversion from only few nanoparticles in real time at realistic reaction conditions and because it (ii) demonstrates the potential of NSM when used in a batch reactor configuration to enable single particle reactivity measurements for arbitrary reactions at technically relevant reaction conditions, provided that batch reactors can be created in a controlled fashion. We also note that we cannot present the TOF here because the exact number of particles trapped in the batch reactor is not known.

*Conclusions*

We have demonstrated the use of nanofluidic scattering microscopy, NSM, to measure concentration changes inside nanofluidic systems based on the methods' high sensitivity to RI changes, which are reflected in a reduced light scattering intensity from a nanochannel if the RI difference between liquid and channel wall material is reduced. Applying this concept to the scrutiny of $H_2O_2$ diffusion into a water filled nanochannel, we have verified the validity of the macroscopic description of molecular diffusion in our nanochannels by extracting a bulk $H_2O_2$ diffusion coefficient in water that is in excellent agreement with the literature value[42] of $D = 2 \cdot 10^{-9} \ m^2/s$. These results thus advertise NSM for experimental studies of diffusion in (even more) nanoconfined systems where molecular interactions with the nanochannel walls may become sizable and the dominant contribution to diffusive molecular transport.

As a second key result, we have demonstrated that NSM enables the visualization, tracking and counting of optically dark metal nanoparticles, such as Pt that is optically lossy in the visible spectral range due to widely abundant interband transitions. This is a significant step because investigations of strongly absorbing metal nanoparticles, as well as plasmonic systems like Au in the sub-50 nm particle size regime, to date are not possible by traditional dark-field scattering microscopy. Hence, NSM can expand the nanoparticle size and composition range accessible with this type of optical microscopy widely used for single nanoparticle studies.

As the third key result on the example of the catalytic $H_2O_2$ decomposition over Pt, we have introduced NSM as a tool for single particle catalysis that enables *in situ* measurements of absolute reactant concentrations directly adjacent to a single active catalyst nanoparticle and along the nanochannel that hosts the particle. Therefore, NSM can resolve the presence or absence of concentration gradients induced by reactant conversion over single nanoparticles and thereby shed light on the interplay between mass transport and surface reaction governed catalyst activity in nanoconfined reaction environments, as well as provide a direct and quantitative measure of reactant consumption, as illustrated for the spontaneous batch reactor. Furthermore, by simultaneously tracking the rate of $O_2$ bubble growth in the nanochannel, we have been able to derive single nanoparticle ToFs that reflected the wide size distribution of the Pt nanoparticles used.

In a wider perspective, our results advertise NSM as a versatile optical microscopy method that, once fully developed, has the potential to enable single particle reactivity measurements for arbitrary reactions without fluorescent labels or other enhancement mechanisms at technically relevant reaction conditions, and in nanoconfinement that mimics porous catalysts support materials.

**Methods**

*Instruments:* The dark-field nanochannel scattering microscopy (NSM) experiments were carried out on a Zeiss Axio Observer Z1 microscope equipped with a Thorlabs Solis-3C LED light source. A Zeiss 50x dark-field objective together with a dark-field reflector cube was used. The scattered light was recorded with an Andor iXon Ultra 888 EMCCD camera, set to take kinetic series of 1000 pictures each with an exposure time of 0.1 s. The transmission electron microscopy (TEM) images were taken with a FEI Tecnai T20 that was operated at 200 kV and had a LaB6 filament, as well as a Orius CCD camera installed. The SEM imaging of the platinum particles and the open nanochannels was carried out on a Zeiss Supra 60 VP with a secondary in-lens electron detector operated at 15 kV. Working distance was set to 2.5 mm.

*Data Evaluation:* The images from the camera were recorded using the Andor Solis Software and saved as full-scale tiff-images. A custom LabView program was used to read in the images and separate the channels, which were then evaluated individually. The two main functions of the program are the counting of pixels whose count values surpass a certain threshold (bubble extension) and averaging the scattering from each side of the nanochannel (concentration determination), both of which were recorded as function of time. The bubble-extension-over-time data was smoothed by fitting a cubic spline fit in LabView evaluation program, and a derivative was taken to determine the bubble extension speed. The measurement of the particle diameter was done in ImageJ after calibrating the size of the SEM picture with the provided scale bar.

*Reagents and Nanoparticles:* All solutions were prepared using ultrapure water (Milli-Q IQ 7000 water purification, Merck). The same water was used when flushing and cleaning the nano/microfluidic system. Hydrogen peroxide was bought from Sigma-Aldrich ($H_2O_2$, 35% w/w in $H_2O$). Platinum nanoparticles from nanoComposix (PTCB70-10M, BioPure Platinum Nanoparticles – Bare (Citrate), 70 nm) were acquired.

*Fluidic chip fabrication:* The nanofluidic systems were fabricated by etching all fluidic structures (nanochannels, microchannels, and inlets) into thermally oxidized silicon substrates and bonding thin glass lids to the structured substrates for sealing, as described by Levin et al.[22,23]. The nanofabrication can be summarized as follows: First 4" (100) silicon wafers were cleaned with Standard Clean 1, 2% HF Dip and Standard Clean 2. The cleaned wafers were wet oxidized at 1050° C to an oxide layer thickness of 2000 nm. Nanochannels were etched into the thermally grown oxide with fluorine-based reactive ion etching (RIE) using a Cr hard mask patterned with e-beam lithography and chlorine-based RIE. To create vertical constrictions within the nanochannels, they were first etched to the nominal constriction depth, then photoresist lines were patterned using laser lithography on the hardmasks across the nanochannels at the constriction positions, and then the nanochannels were etched to their final depth. Subsequently, all microfluidic structures were etched into the surface by RIE using photoresist etch masks, typically patterned by direct laser lithography, and inlets were fabricated with deep reactive ion etching. Finally, the substrates were cleaned with Standard Clean 1 along with 175 µm thick 4" Borofloat33 glass wafers, and the surfaces of both the substrates and the glass wafers were plasma treated with O2 plasma (50W, 250mTorr) to allow the glass lids to be pre-bond to the substrates prior to fusion bonding them (550 °C, 5 hours). The bonded wafers were then cut into individual fluidic chips.

**Supporting Information**

This material is available free of charge via the Internet at http://pubs.acs.org.

Calculated $H_2O_2$ diffusion profiles, collection of Pt nanoparticle TEM images, light scattering intensity changes for particles in nanochannels, Pt particle counting and bubble evaluation for Scenario I and II, images showing Pt particles being placed not at the trap, nanochannel flow speed assessment during bubble formation, estimation of how much of a nanochannel is occupied by an $O_2$ bubble, BES trace start and end conditions, $O_2$ bubble formation stills for corresponding movie, estimation of time delay between the arrival of $H_2O_2$ in the up and downstream sections of the nanochannel, atomistic representation of a Pt nanoparticle for active site assessment, theoretical calculation of nanochannel scattering cross sections, estimation of the influence of oxygen solubility, turnover frequency calculation from bubble expansion speed. (PDF)

Video SV1: NSM recording of Pt nanoparticles moving in nanochannels under the influence of Brownian motion and convective flow. (AVI)

Video SV2: NSM recording of rapid successive bubble formation (Scenario I), followed by upstream bubble formation (Scenario II). (AVI)

Video SV3: NSM recording of $O_2$ bubble formation in a nanochannel after the arrival of $H_2O_2$ at a single particle. (AVI)

Video SV4: NSM recording of $O_2$ bubble evolution against the convective flow with particles being pushed out by the bubble. (AVI)

Video SV5: NSM recording of the evolution of two $O_2$ bubbles, whereof one is transiently stuck in the nanochannel. (AVI)

Video SV6: NSM recording of the spontaneous creation of a "batch" reactor, where a section of the nanochannel is enclosed by $O_2$ bubbles. (AVI)

**Corresponding Author**

clangham@chalmers.se

**Author Contributions**

The manuscript was written through contributions of all authors. All authors have given approval to the final version of the manuscript.

**Acknowledgements**

This research has received funding from the Swedish Research Council (VR) Consolidator Grant project 2018-00329 and the European Research Council (ERC) under the European Union's Horizon Europe research and innovation program (101043480/NACAREI). Part of this work was carried out at the Chalmers MC2 cleanroom facility and at the Chalmers Materials Analysis Laboratory (CMAL).

**TOC Figure**

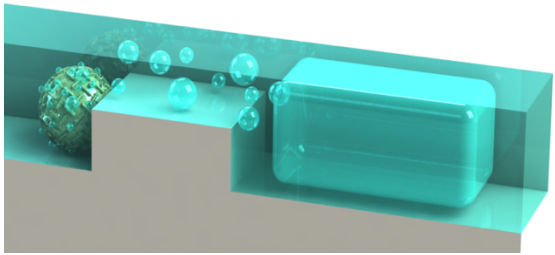

# Supplementary Material

# For

Label-free Imaging of Catalytic $H_2O_2$ Decomposition on Single

Colloidal Pt Nanoparticles using Nanofluidic Scattering Microscopy

*Björn Altenburger[1], Carl Andersson[1], Sune Levin[2], Fredrik Westerlund[2], Joachim Fritzsche[1]*

*and Christoph Langhammer[1]\**

[1]Department of Physics, Chalmers University of Technology; SE-412 96 Gothenburg,

Sweden

[2]Department of Biology and Biological Engineering, Chalmers University of Technology;

SE-412 96 Gothenburg, Sweden

*Corresponding author: clangham@chalmers.se

**Section I: Supplementary figures**

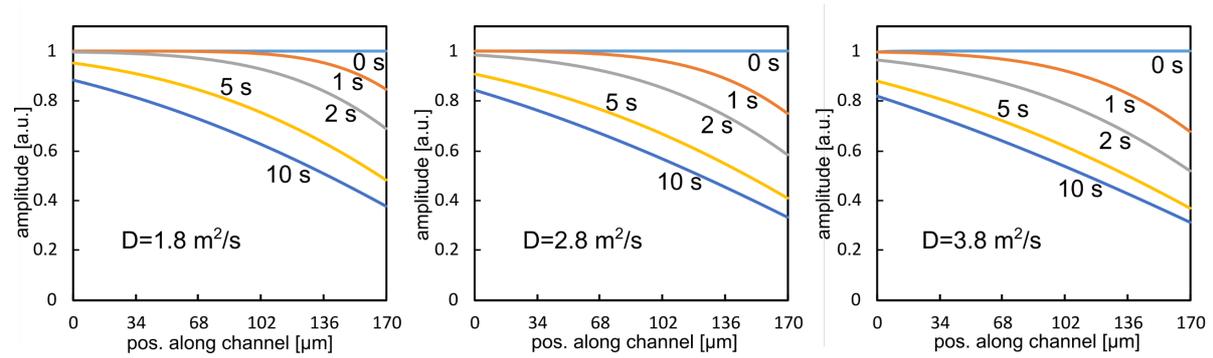

**Figure S1.** Calculated time evolution of $H_2O_2$ concentration profiles in a quadratic nanochannel with 150 nm side length for bulk diffusion constants D = 1.8 $m^2$/s, D = 2.8 $m^2$/s and D = 3.8 $m^2$/s.

a) Before H₂O₂ exposure

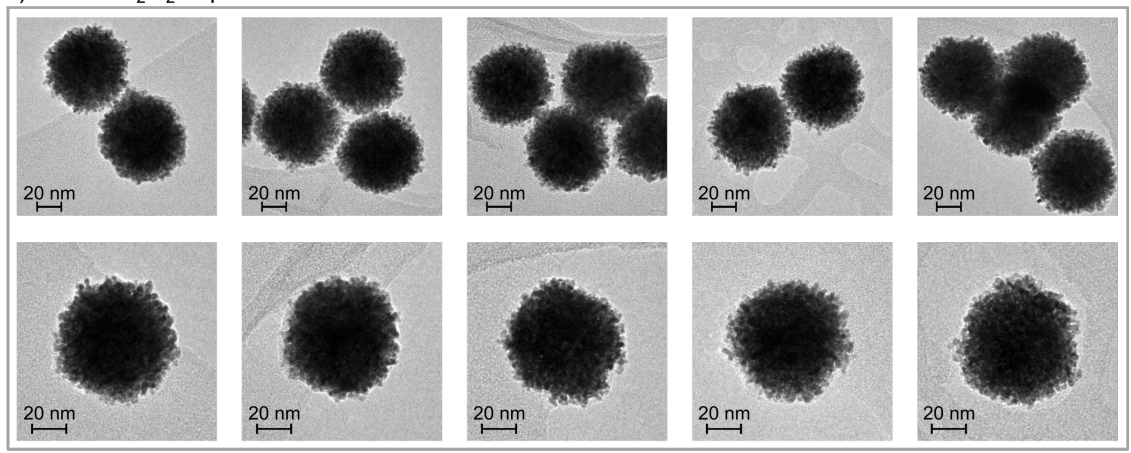

b) After H₂O₂ exposure in solution

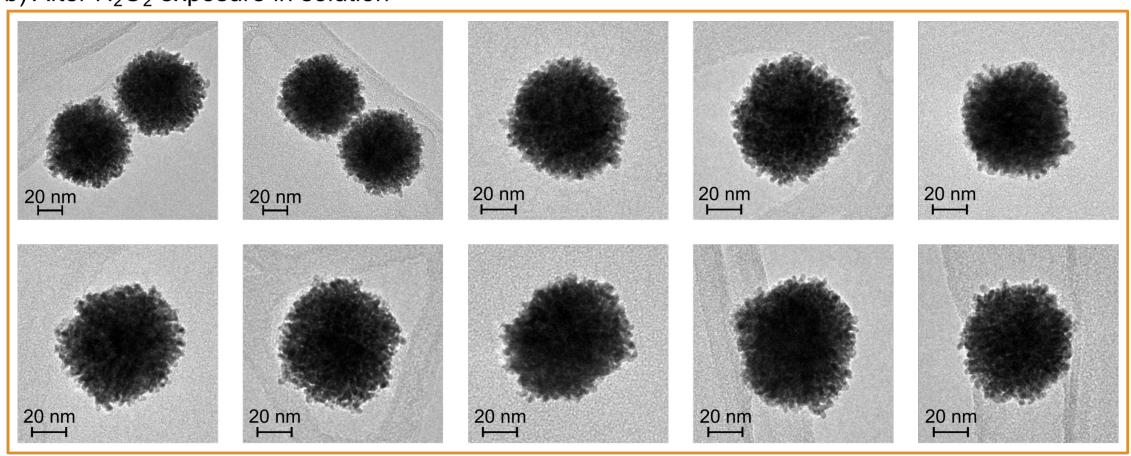

c) After H₂O₂ exposure on carbon substrate

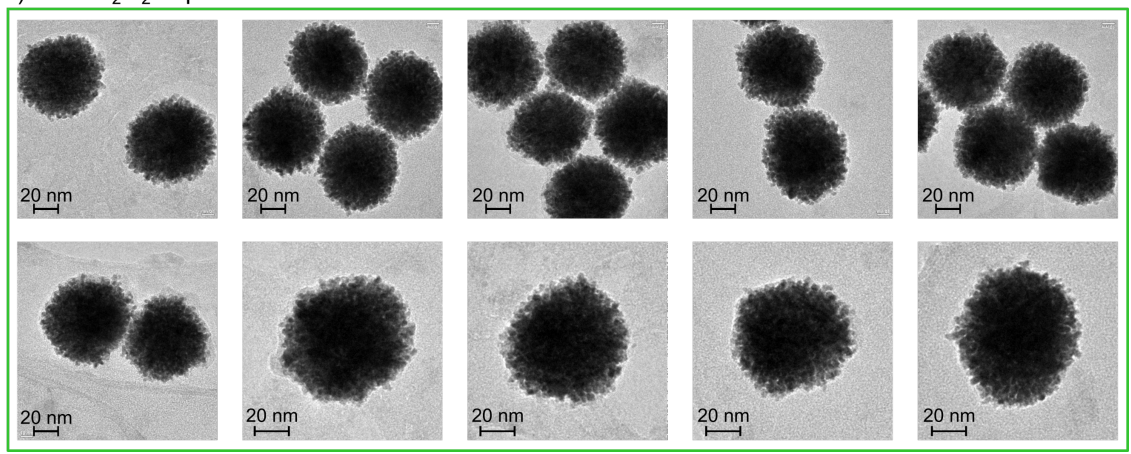

**Figure S2**. Selection of TEM-pictures of the colloidal Pt particles used in our experiments before and after exposure to reactions. All images were obtained by drop casting the particles onto holey carbon TEM grids. a) Pt particles as extracted from the storage solution before they have been in contact with 30% $H_2O_2$ in water. b) Platinum particles after exposure to $H_2O_2$ decomposition reaction conditions in aqueous solution for ca. 5 minutes. c) Platinum particles that have been subjected to $H_2O_2$ after drop casting them onto holey carbon. While the particles seem quite similar, it is evident that they are not identical and vary in surface structure and size. There is also no effect by $H_2O_2$ on the particles visible, only the carbon support seems rougher in the TEM sample that was treated with $H_2O_2$.

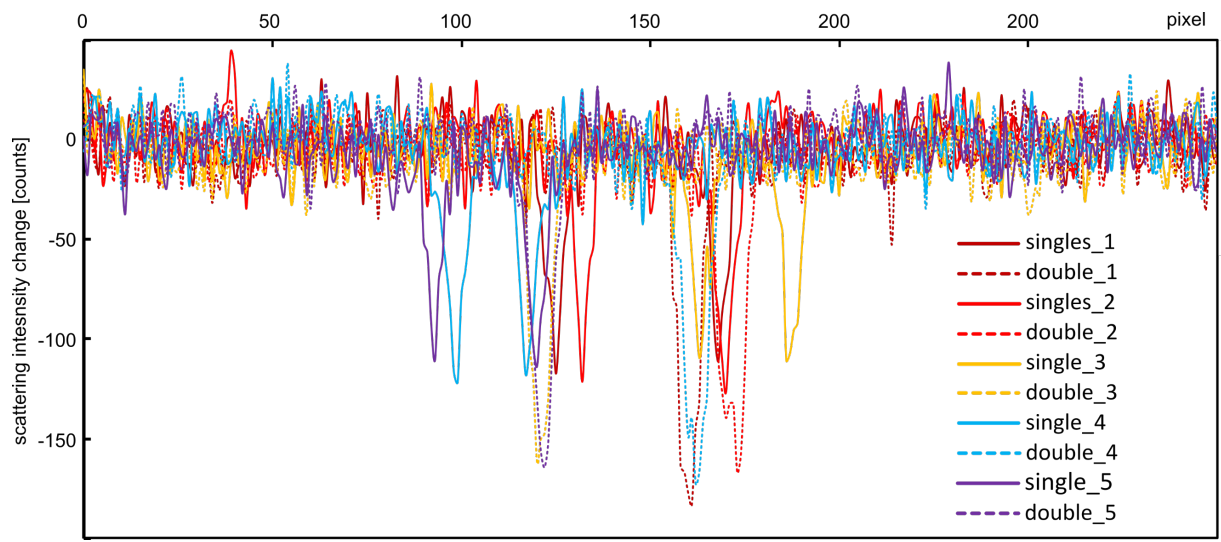

**Figure S3.** Experimental light scattering intensity changes measured along the channel direction (expressed as pixel-# along the x-axis) for nanochannels containing two diffusing Pt particles that transiently combine in a diffraction limited spot and from a "dimer". The color-coded curves show the scattering intensity difference between channel and particles in their separated and "dimer" state for each channel. They reveal a decrease in intensity of about 40 % for the dimers compared to their corresponding single particles.

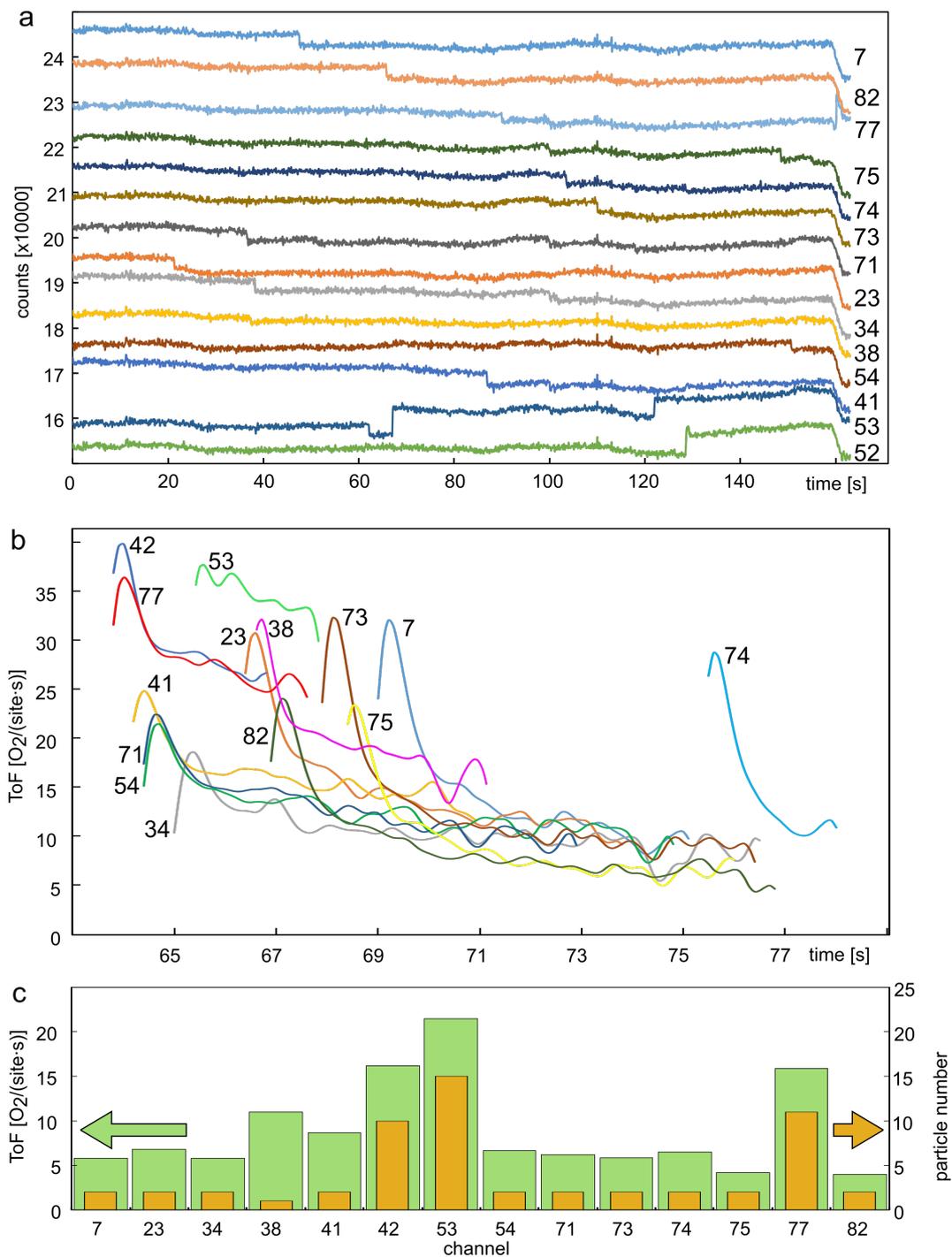

**Figure S4.** Pt particle counting and bubble evaluation. a) Integrated scattering intensity time traces recorded for the trap area for the measurement shown in **Figure 4** in the main text for all channels that show bubble formation according to Scenario I. The arrival of individual particles can be seen as distinct steps to lower intensity. However, it can also be seen that when multiple particles accumulate, the scattering intensity at the trap instead increases and becomes higher than in the empty reference channels numbered 42, 53 and 77. B) ToF traces for the channels shown in a) for an $H_2O_2$ concentration of 30% in water. c) Comparison of the counted (and for the largest numbers estimated) number of particles in each channel and the corresponding ToF derived from the BES.

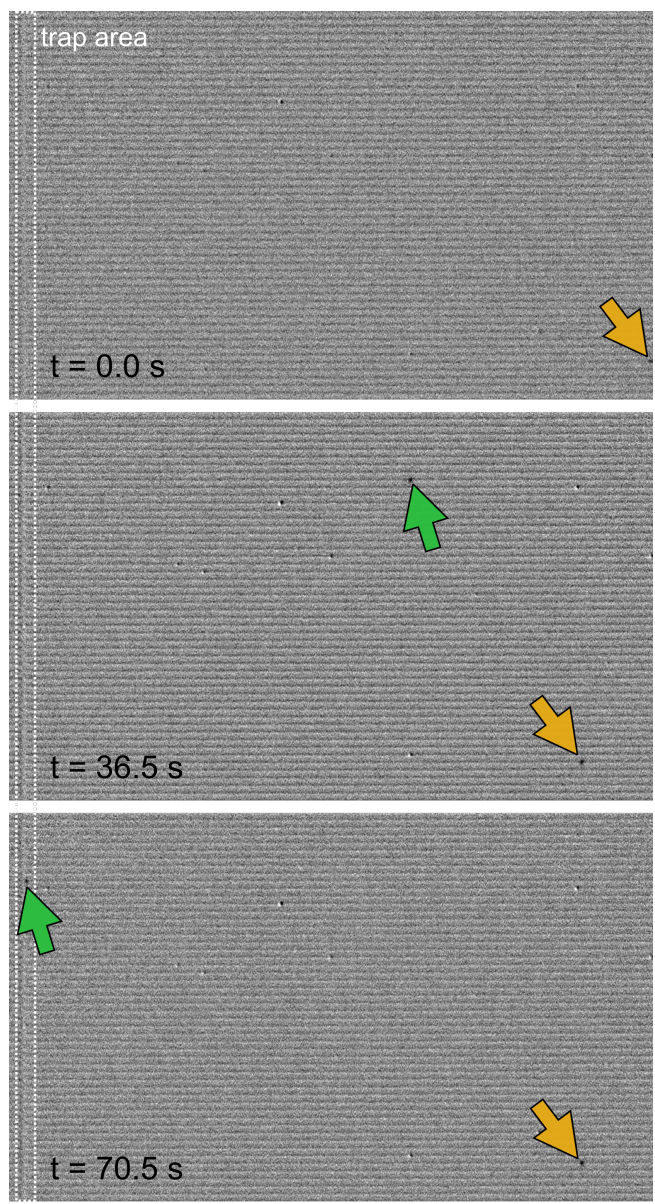

**Figure S5.** Citrate coated Pt particles getting stuck in nanochannels before the trap. The particle marked with orange arrow enters the field of view at t = 0s as it is pushed through the channel but gets stuck due to electrostatic interaction with the wall at t = 36.5 s and remains in the same position at t = 70. 5 s. The particle marked with a green arrow gets transiently stuck at 36.5 s before eventually reaching the trap at 70.5 s.

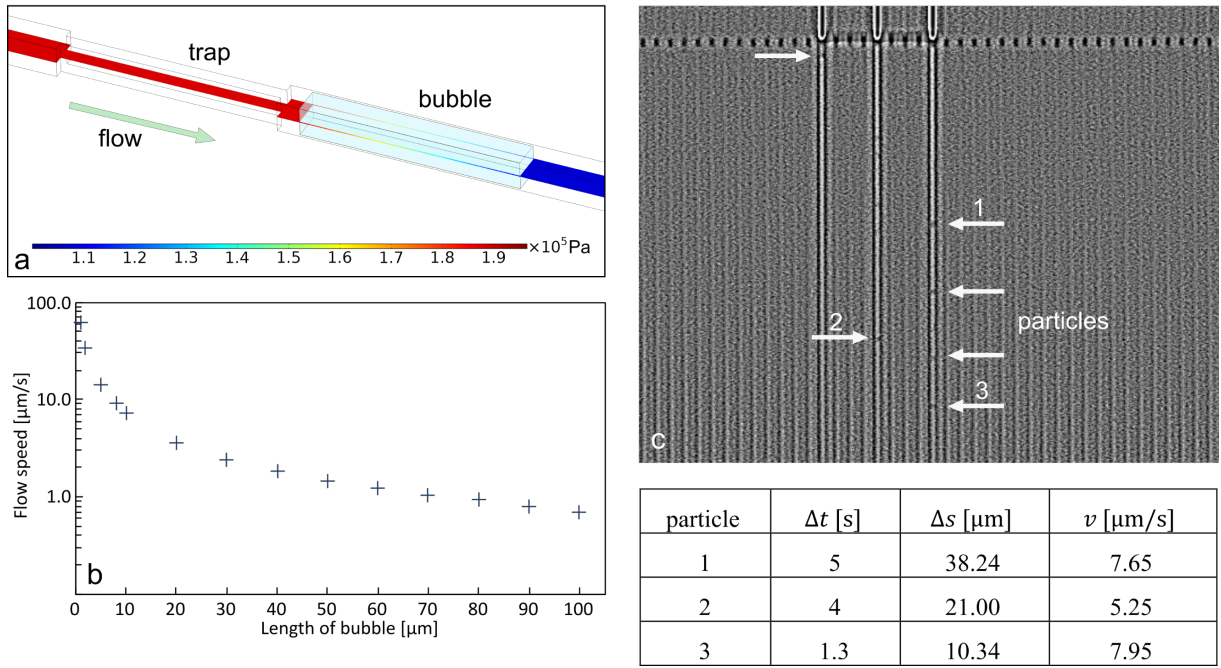

| particle | $\Delta t$ [s] | $\Delta s$ [µm] | $v$ [µm/s] |
|----------|----------------|-----------------|------------|
| 1        | 5              | 38.24           | 7.65       |
| 2        | 4              | 21.00           | 5.25       |
| 3        | 1.3            | 10.34           | 7.95       |

**Figure S6.** Water flow speed assessment for the nanochannels at 2 bar inlet pressure. a) Comsol simulation of the pressure drop at the position of a 1 µm long bubble that occupies 75.5% of a 150 nm x 150 nm nanochannel and that is positioned after the constriction that traps the nanoparticles (slice through the middle shows pressure as color gradient). While the pressure remains constant throughout the trap, it is decreased by 1 bar after the bubble. b) Simulated flow speed inside the nanochannel as a function of the length of a bubble that is growing inside that channel. The maximum length of the bubble is 170 µm, which corresponds to the length of the nanochannel after the constriction. c) Measurement of the flow speed inside nanochannels of identical dimensions as in the simulation obtained by tracking of the movement of Pt nanoparticles (arrows) induced by the flow. The differential dark-field image shows how particles can be seen moving out of the channels when the pressure at the inlets of the chip is inverted (flow from top to bottom). The three channels that stand out are filled with a bubble on the other side of the trap (top). Tracking several particles while the same pressure is applied as during bubble formation allows an estimation of the flow speed when a bubble is filling the channels. The table lists the correspondingly obtained results from three different Pt particles.

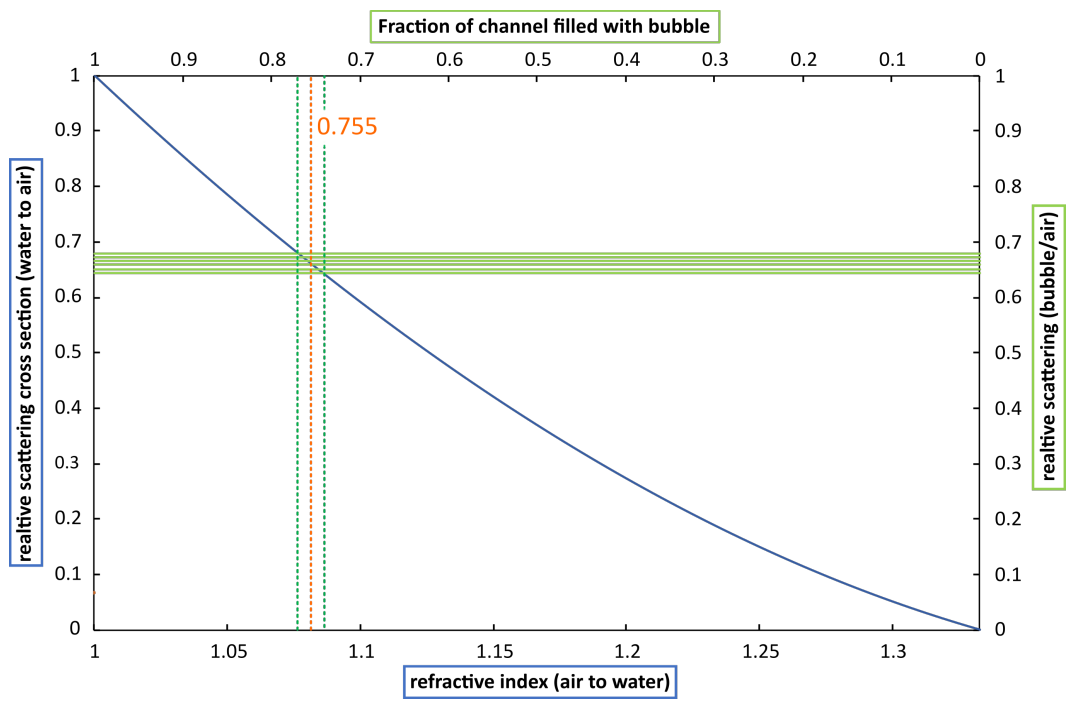

**Figure S7.** Comparison to estimate the fraction of the nanochannel that is filled with gas when a bubble is filling the channel. Blue x- and y-axis: Calculated relative scattering cross section between water and air (as a mimic of $O_2$ in the gas phase) plotted as a function of refractive index systematically changing from air to water. The relative scattering cross section of a completely air-filled channel has been set to 1 and for a completely water filled channel it is set to 0. Green x- and y-axis: Measured scattering from a set of $O_2$ bubble-filled channels scaled relative to the scattering of the same channels filled with air using the same scaling as for the simulated data. The comparison of the experimental with the theoretical curve yields the estimate that the gas bubble occupies the channel volume to ca. 75.5%.

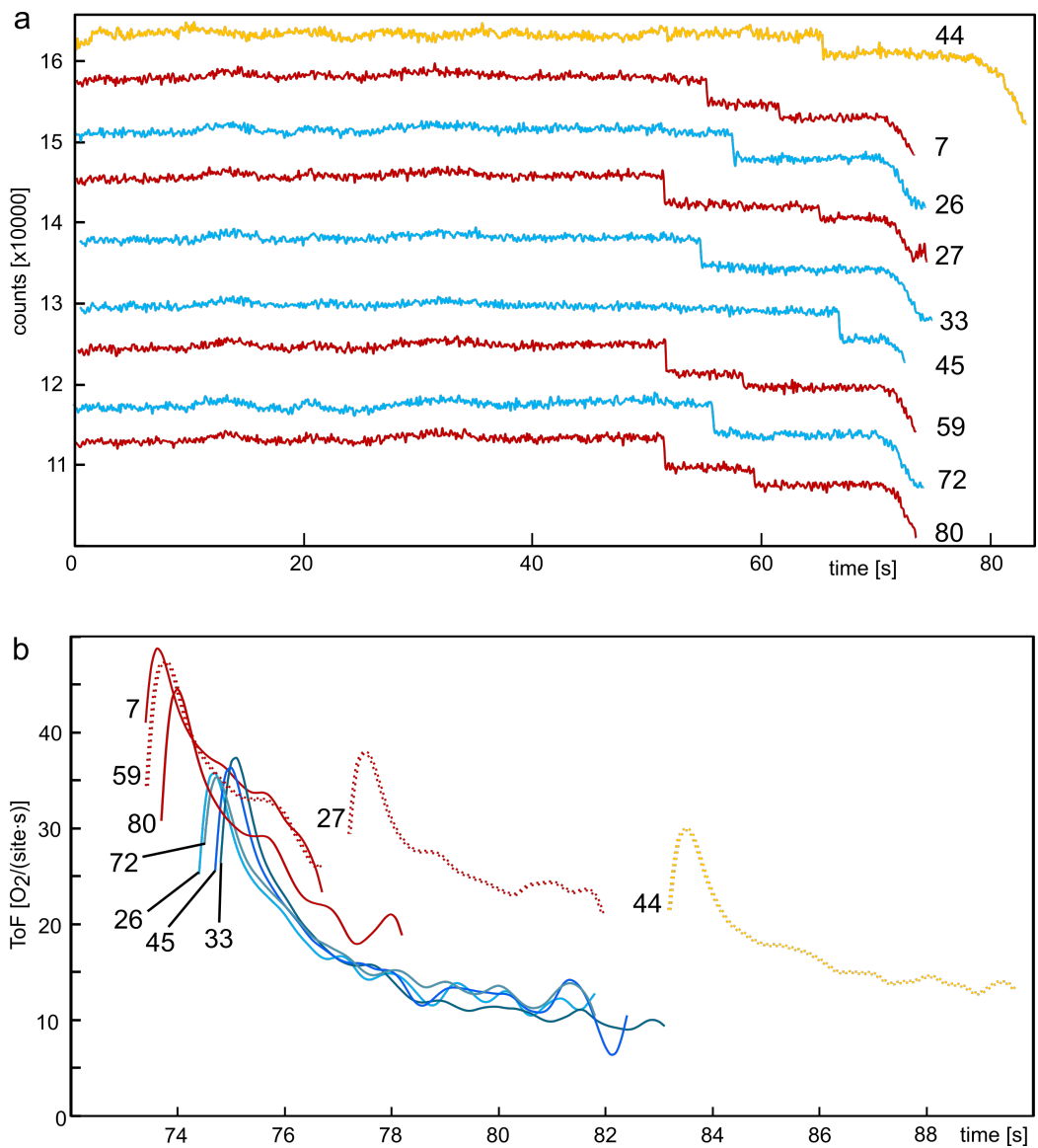

**Figure S8.** a) Integrated scattering intensity time traces of the trap area for the measurement shown in **Figure 5** in the main text for all channels that show bubble formation according to Scenario II. The arrival of single or multiple particles can be seen as steps in the intensity time trace. Red curves represent channels with two particles getting trapped (seen as two individual steps), blue and yellow curves represent channels where a single particle is trapped. b) ToF traces calculated from the BES for Scenario II bubbles forming in the same channels as analyzed for trapped particles in panel a). Solid ToF traces mark cases where the particle(s) remained stationary at the trap, whereas dashed ToF traces mark cases where particles were pushed out by the developing bubble. The color code is the same as in a). Blue and red ToF traces were measured for a 15% $H_2O_2$ reactant solution and the yellow trace was measured for a 30% $H_2O_2$ reactant solution.

a) Start

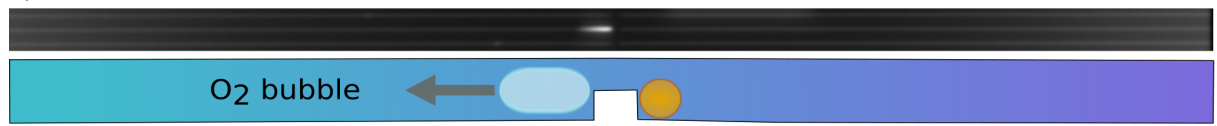

b) End

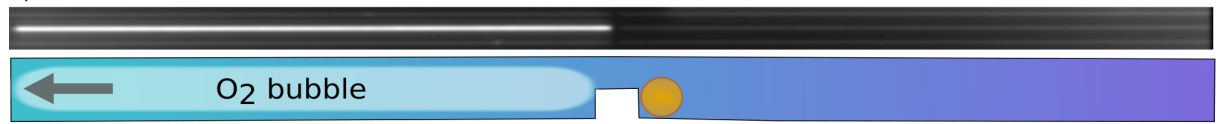

c) Detachment

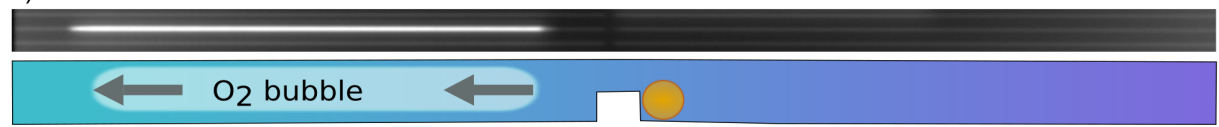

| 0 | pixels | 128 | 256 | 384 | 512 | 640 | 768 | 896 | 1024 |

**Figure S9**. Start and end points of a BES trace. A) The BES trace starts as soon as a visible bubble starts to form, that is, when the brightness of some pixels along the nanochannel is reaching a certain threshold. The BES trace then corresponds to the number of pixels over this threshold over time. b) The tracing of the bubble extension speed is ended when the bubble reaches the edge of the field of view of the camera, since from this point forward the number of bright pixels remains constant. c) In some cases, the bubble detaches from the trap and moves through the nanochannel. This will also be regarded as an end to the BES trace since the bubble is not growing anymore and since the connection to the particle is lost.

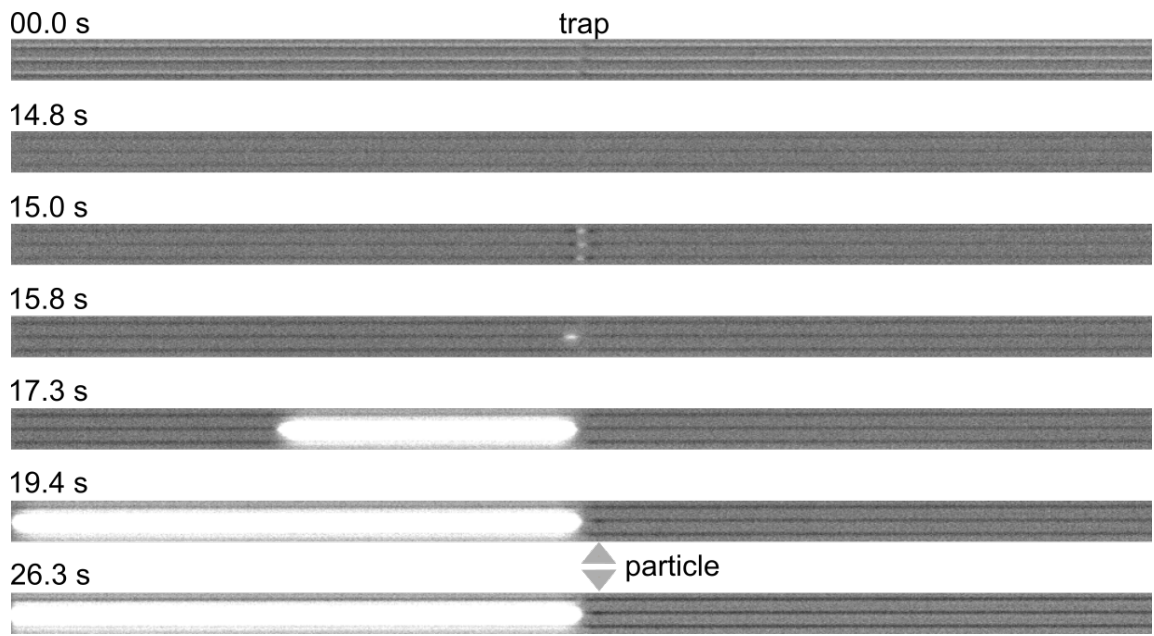

**Figure S10. Still images of characteristic time points in movie SV3.** For this movie, a frame obtained at 14.5 s (immediately before $H_2O_2$ entered the channel) has been subtracted from all subsequent frames to highlight subsequent changes in scattering intensity. 0.00 s) Water-filled channels with the trap area in the middle. The particle is already in place in the nanochannel in the center. 14.8 s) The 30% $H_2O_2$ solution is starting to fill the channel when the flow of water was switched off. This is visible by the darker channels and the overall slightly darker right side of the image. 15.0 s) Disturbances cause movement of the whole chip, which makes the trap area shortly visible in these differential images. Reasons for these disturbances can be the pressure switch from the water inlet side to the $H_2O_2$ inlet side of the chip or the touching of the pressure switches which transfers to the chip via the pressure pipes. 15.8 s) The $O_2$ bubble starts to form on the side of the trap that is opposite the particle in the center channel after the $H_2O_2$ solution has been flushed in. 17.3 s) The bubble has extended about half the way to the edge of the field of view. 19.4 s) The bubble has reached the edge of the field of view, such that the BES trace is terminated. Since the convective flow is now largely stopped by the bubble blocking the channel, the particle drifts a short distance away from the trap (arrows). 26.3 s) The bubble has extended beyond the field of view. On the particle side, the center channel and its neighboring channels have the same low brightness, indicating so that they are filled with about the same concentration of $H_2O_2$.

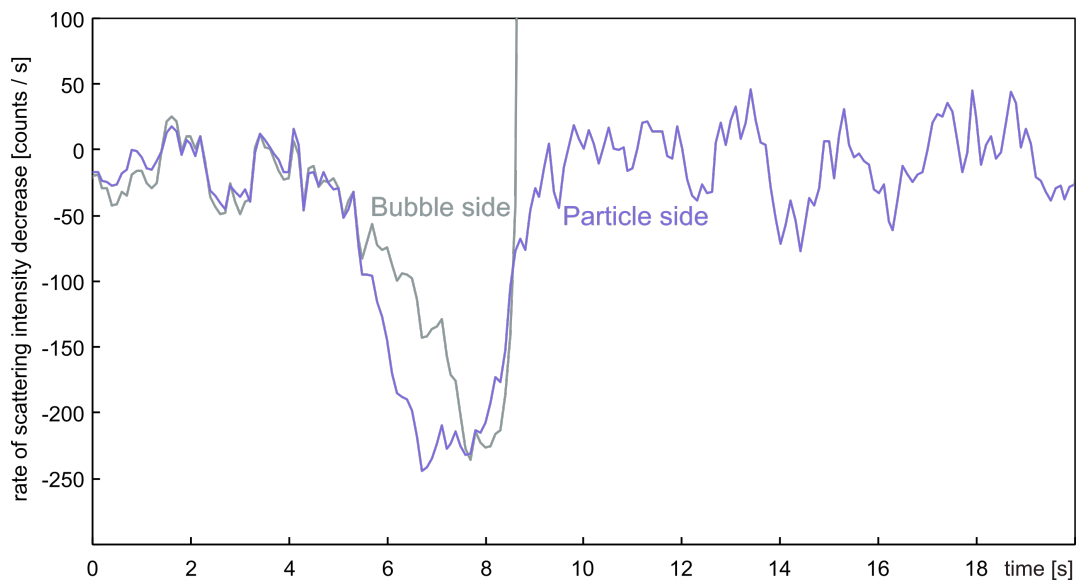

**Figure S11.** Time derivative of the channel scattering intensity traces shown in Figure 5a in the main text that reveals a clear shift of the peak of the derivative by about 1 s between the two separately imaged channel sections down and upstream of the particle, respectively. This indicates a time delay of about 1 second for the $H_2O_2$ front reaching these two channel sections.

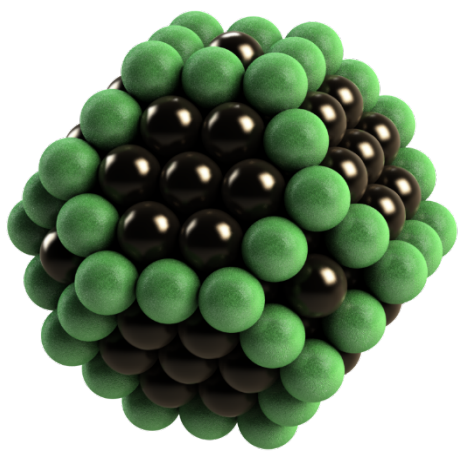

**Figure S12.** Atomistic representation of a Pt nanoparticle with a total of 201 atoms. In this configuration, there are 124 atoms located on the surface, whereof 60 are low-coordinated edge or corner sites, while the remaining 64 atoms are located in terraces. The fraction of edge/corner sites at the surface is thus 48.4%. Even though the size of this particle is with only 1.7 nm diameter[1] at the lower end of the observed crystallite sizes of the Pt particles used in our experiments (see **Figure 3a** in the main text), it rationalizes the edge & corner/terrace site ratio of ca. 50 % we use in our discussion as reasonable and representative for the Pt nanoparticles at hand

**Section II: Supplementary Derivations**

*1.  Theoretical calculation of nanochannel scattering cross sections*

The derivation of the scattering cross section formula for a nanochannel, **Error! Reference source not found.**in the main text, is described in detail in chapter 8 in the book by Bohren and Huffman[2]. For convenience, we summarize the key steps below.

To mathematically describe the rectangular channels from the experiment, we use a cylinder of infinite length, whose diameter corresponds to the cross section of the experimental nanochannel. Accordingly, the derivation uses cylindrical coordinates as given in **Figure S10**.

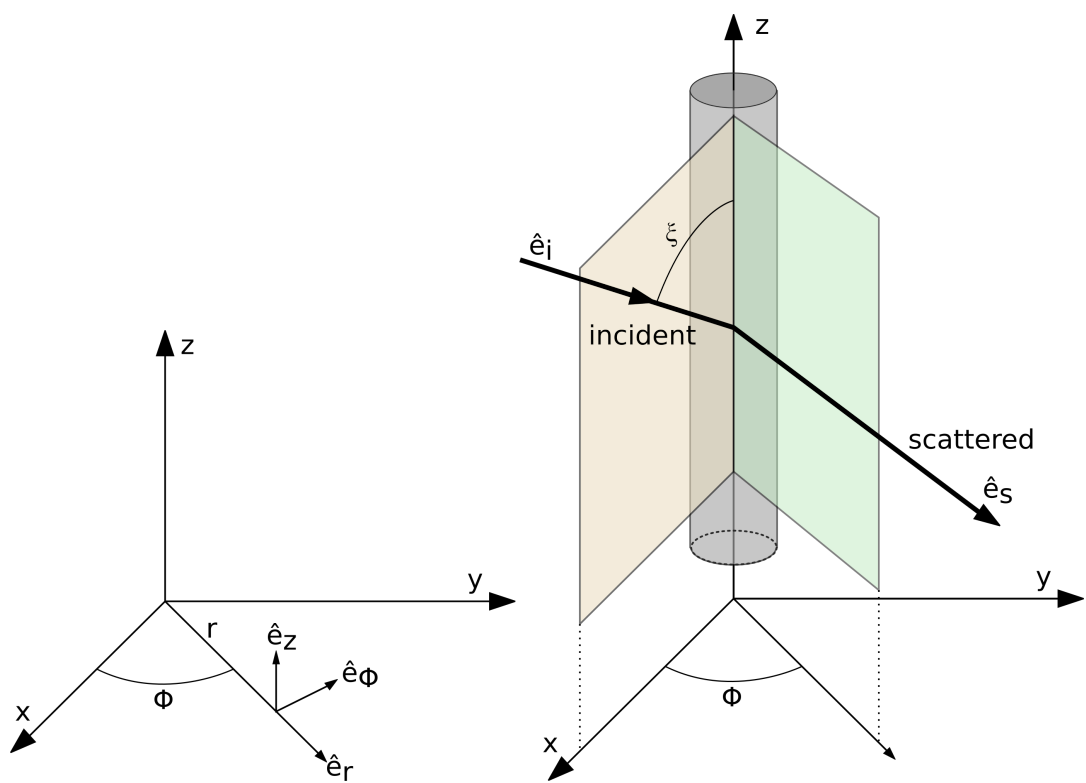

**Figure S13.** Cylindric coordinate system and a section of the infinite cylinder with the Poynting vector of the incident and scattered light.

The function that solves the wave equation in cylindrical coordinates can be written in a form that separates into a radial, angular and vertical part, where $h = -k \cos \zeta$ is the separation constant and $\rho = r\sqrt{k^2 - h^2}$ with $k$ as wavenumber.

$$\psi_n(r, \phi, z) = Z_n(\rho)e^{in\phi}e^{ihz} \quad (n = 0, \pm 1, \dots)$$

Linearly independent solutions to $Z_n$ are the Bessel functions of the first and second kind, $J_n$ and $Y_n$ respectively, where $n$ denotes their integral order. The harmonic functions generated from the equation above,

$$M_n = \nabla \times (\hat{e}_z \psi_n), \qquad N_n = \frac{\nabla \times M_n}{k}$$

will be used to express the incident electric field of a plane wave $E_i = E_0 e^{ik\hat{e}_i x}$ that is incident onto a cylinder of radius $a$ in the direction of $\hat{e}_i = -\sin\zeta \hat{e}_x - \cos\zeta \hat{e}_z$ with $\zeta$ being the angle between cylinder axis and the incident wave. For an incident electric wave that is parallel to the xz-plane, the expansion in cylinder harmonics is

$$E_i = \sum_{n=-\infty}^{\infty} [A_n M_n^I + B_n N_n^I].$$

The generating function for the cylinder harmonics in then $J_n(kr\sin\zeta)e^{in\phi}e^{-ikz\cos\zeta}$, as we need to exclude the Bessel functions of the second kind to avoid an infinite electric field at $r = 0$. Following Bohren and Huffman, the coefficients are given as

$$A_n = 0, \qquad B_n = \frac{E_0(-i)^n}{k \sin\zeta}$$

such that for the incident wave, when $E_n = E_0 (-i)^n/k \sin\zeta$,

$$E_i = \sum_{n=-\infty}^{\infty} E_n N_n^I, \qquad H_i = \frac{-ik}{\omega\mu} \sum_{n=-\infty}^{\infty} E_n M_n^I$$

For the internal field, we define first the ration of refractive indexes

$$m = \frac{n_{cylinder}}{n_{surrounding}}.$$

to arrive at

$$E_I = \sum_{n=-\infty}^{\infty} E_n[g_n M_n^I + f_n N_n^I], \qquad H_I = \frac{-ik}{\omega\mu} \sum_{n=-\infty}^{\infty} E_n[g_n N_n^I + f_n M_n^I]$$

For the scattered field, we find

$$E_s = \sum_{n=-\infty}^{\infty} E_n[b_{n1} N_n^3 + ia_{n1} M_n^3], \qquad H_s = \frac{ik}{\omega\mu} \sum_{n=-\infty}^{\infty} E_n[b_{n1} M_n^3 + if a_{n1} N_n^3]$$

Such that we now can describe the Poynting vector

$$S_s = \frac{1}{2}\,\mathrm{Re}(\boldsymbol{E}_s \times \boldsymbol{H}_s^*) \quad S_s = \frac{1}{2}\,\mathrm{Re}(\boldsymbol{E}_i \times \boldsymbol{H}_s^* + \boldsymbol{E}_s \times \boldsymbol{H})_i^*$$

and the absorption, scattering and extinction rates for the incoming light.

$$W_a = -\int_A \boldsymbol{S} \cdot \hat{\boldsymbol{n}}\;dA = W_{ext} - W_s = RL\int_0^{2\pi}(\boldsymbol{S}_{ext})_r\;d\phi - RL\int_0^{2\pi}(\boldsymbol{S}_s)_r\;d\phi$$

Comparing these rates with the projected cross section of the channel, we arrive at the scattering efficiency (for parallel incident light).

$$Q_{sca,p} = \frac{W_s}{2aLI_i} = \frac{2}{x}\left[|b_0|^2 + 2\sum_{n=1}^{\infty}(|b_n|^2 + |a_n|^2)\right]$$

The coefficients $a_n$ and $b_n$ can be found in the book by Bohren and Huffman[2], and we arrive at the scattering efficiencies for both parallel, and, in similar manner, orthogonally polarized incident light.

$$Q_{sca,p} = \frac{\pi^2 x^3}{8}(m^2-1)^2 = \frac{\pi^2 k^3 a^3}{8}(m^2-1)^2$$

$$Q_{sca,o} = \frac{\pi^2 x^3}{4}\left(\frac{m^2-1}{m^2+1}\right)^2 = \frac{\pi^2 k^3 a^3}{4}\left(\frac{m^2-1}{m^2+1}\right)^2$$

From those, we use $x = ka = 2\pi a/\lambda$ and as geometrical cross section $A_\emptyset = \pi a^2$ to arrive at the scattering cross sections for both polarization directions of incident light, as well as for unpolarized light, which corresponds to **Error! Reference source not found.** in the main text.

$$\sigma_{sca,p} = \frac{A_\emptyset^2 k^3 L}{4}(m^2-1)^2$$

$$\sigma_{sca,o} = \frac{A_\emptyset^2 k^3 L}{2}\left(\frac{m^2-1}{m^2+1}\right)^2$$

$$\sigma_{sca,u} = \frac{A_\emptyset^2 k^3 L}{4}(m^2-1)^2\left(\frac{1}{2} + \frac{1}{(m^2+1)^2}\right)$$

## 2. *Influence of oxygen solubility*

For the estimation of the influence of the $O_2$ solubility in water on the bubble formation in a nanochannel, we start at an average BES of 6.3 µm/s, which can be translated into an $O_2$ production rate of $2.5 \cdot 10^6 \; O_2/s$ per particle when using **Error! Reference source not found.** in the main text.

From Figure 2d in the main text, we can estimate the flow speed of the liquid in the channel at an inlet pressure of 2 bar to be 40 µm/s. Multiplied with the channel cross section of 150 x 150 nm$^2$, we arrive at a volume flow of $7.84 \cdot 10^{-19}\; m^3/s$. Using now the solubility of $1.22\; mol/m^3$ and Avogadro's constant, we see that about $0.6 \cdot 10^6\; O_2/s$ could be transported away by the flow in the channel, which is about a quarter of the above estimated $O_2$ production rate.

This will however be only the case at the very start of the reaction, because as soon as a bubble has formed, the flow speed will be drastically decreased (see also **Figure S6**), thereby limiting the transport of dissolved oxygen and thus significantly reducing this effect.

### 3. *Turnover frequency from bubble expansion speed*

At the start, we need to derive the amount of $O_2$ produced per second. To do this, we assume that the bubble is rectangular as the channel and has a cross section that is the geometrical cross section of the channel times the filling factor determined in **Figure S7**.

$$A_{bub} = 0.755 A_{channel}$$

Multiplied by the BES, which has the unit of m/s, we arrive at the bubble volume expansion speed in m$^3$/s. Divided by the volume of 1 mol of $O_2$ gas, $V_{O2}^{mol} = 22.4\ l/mol$, we get the number of moles of molecular oxygen produced within 1 s.

$$R_{o2} = \frac{A_{bub}\ BES}{V_{O2}^{mol}}$$

Multiplying this with Avogadro's number, we get the absolute amount of $O_2$ molecules produced per second. Divided by the number of active sites $N$, we get the number of $O_2$ molecules produced per site and second, which corresponds to the turnover frequency.

$$ToF = \frac{0.755\ A_{channel}\ BES\ N_A}{V_{O2}^{mol} N}$$